\documentclass[12pt]{article}
\usepackage{amssymb}
\usepackage{amsmath}
\usepackage{graphicx}
\usepackage{indentfirst}
\usepackage{cite}
\usepackage{appendix}

\allowdisplaybreaks

\begin{document}

\title{Cross sections for 2-to-1 meson-meson scattering}
\author{Wan-Xia Li$^1$, Xiao-Ming Xu$^1$, and H. J. Weber$^2$}
\date{}
\maketitle \vspace{-1cm}
\centerline{$^1$Department of Physics, Shanghai University, Baoshan,
Shanghai 200444, China}
\centerline{$^2$Department of Physics, University of Virginia, Charlottesville,
VA 22904, USA}

\begin{abstract}
We study the processes $K\bar{K} \to \phi$, $\pi D \to D^\ast$, 
$\pi \bar{D} \to \bar{D}^\ast$, and the production of $\psi (3770)$, 
$\psi (4040)$, $\psi (4160)$, and
$\psi (4415)$ mesons in collisions of charmed mesons or charmed strange mesons.
The process of 2-to-1 meson-meson scattering involves a quark 
and an antiquark from the two initial mesons annihilating into a gluon and
subsequently the gluon being absorbed by the spectator quark or antiquark. 
Transition amplitudes for the scattering process derive from the transition 
potential in conjunction with mesonic quark-antiquark wave functions and the 
relative-motion wave function of the two initial mesons. We derive these 
transition
amplitudes in the partial wave expansion of the relative-motion wave function
of the two initial mesons so that parity and
total-angular-momentum conservation are maintained. We calculate flavor and 
spin matrix elements in accordance with the transition potential and
unpolarized cross sections for the reactions using the transition
amplitudes. Cross sections for the production of $\psi (4040)$, $\psi (4160)$,
and $\psi (4415)$ relate to nodes in their radial wave functions.
We suggest the production of $\psi (4040)$,
$\psi (4160)$, and $\psi (4415)$ as probes of hadronic matter that
results from the quark-gluon plasma created in ultrarelativistic heavy-ion 
collisions.
\end{abstract}



\vspace{0.5cm}
\leftline{\bf I. INTRODUCTION}
\vspace{0.5cm}

Elastic meson-meson scattering produces many resonances. 
Starting from meson-meson scattering amplitudes obtained in
chiral perturbation theory~\cite{CPT}, elastic scattering has been
studied within nonperturbative schemes, for example, 
the inverse amplitude method~\cite{Hannah} and
the coupled-channel unitary approaches~\cite{OO}. 
Elastic meson-meson scattering has also been studied with quark 
interchange in the first Born approximation in Ref. \cite{BS} and with
quark-antiquark annihilation and creation in Ref. \cite{SXW}. 
Elastic scattering reported in the literature includes $\pi \pi$
\cite{CPT,Hannah,OO}, $\pi K$ \cite{CPT,Hannah,OO},
$K\bar K$ \cite{OOP,DEW,SBGKLN,AMRR}, $\pi \eta$ 
\cite{OOP,DEW,BFS,GLMOR,BHHKM,BN}, $K \eta$ \cite{OOP},
$\eta \eta$ \cite{SBGKLN}, $\pi \rho$ \cite{LLMP,RO},
$\pi D$ \cite{FN}, $\bar{K} D$ \cite{ICADH,TDKJO}, 
$\bar{K} D^\ast$ \cite{ICADH}, and $D D^\ast$ \cite{ICADH}.
We know that resonances observed in the elastic scattering
are usually produced by a process where two mesons scatter into one meson. 
The 2-to-1 meson-meson scattering includes
$\pi \pi \to \rho$, $\pi \pi \to f_0(980)$, $\pi K \to K^\ast$, 
$K\bar{K} \to \phi$, $\pi \eta \to a_0(980)$, $\pi \rho \to a_1(1260)$,
$\pi D \to D^\ast$, and so on. Since some resonances like $f_0(980)$, 
$a_0(980)$, and $a_1(1260)$ are not quark-antiquark states, we do not study
$\pi \pi \to f_0(980)$, $\pi \eta \to a_0(980)$, $\pi \rho \to a_1(1260)$,
etc. in the present work. The reactions $\pi \pi \to \rho$ and
$\pi K \to K^\ast$ have been studied in Ref. \cite{YXW} via a process
where a quark in an initial meson and an antiquark in another initial meson
annihilate into a gluon and subsequently the gluon is absorbed by the other
antiquark or quark. The resulting cross sections in vacuum
agree with the empirical data.
Since these two reactions also take place in hadronic matter that
is created in ultrarelativistic heavy-ion collisions at the Relativistic Heavy
Ion Collider and at the Large Hadron Collider, the dependence of 
cross sections for the two reactions 
on the temperature of hadronic matter has also been investigated. 
With increasing temperature the cross sections
decrease. In the present work we consider the reactions: 
$K\bar{K} \to \phi$, $\pi D \to D^*$, $\pi \bar{D} \to \bar{D}^*$, 
$D\bar{D} \to \psi (3770)$, $D\bar{D} \to \psi (4040)$,
$D^*\bar{D} \to \psi (4040)$, $D\bar{D}^* \to \psi (4040)$,
$D^*\bar{D}^* \to \psi (4040)$, $D_s^+D_s^- \to \psi (4040)$,
$D\bar{D} \to \psi (4160)$,
$D^*\bar{D} \to \psi (4160)$, $D\bar{D}^* \to \psi (4160)$,
$D^*\bar{D}^* \to \psi (4160)$, $D_s^+D_s^- \to \psi (4160)$,
$D_s^{*+}D_s^- \to \psi (4160)$, $D_s^+D_s^{*-} \to \psi (4160)$,
$D\bar{D} \to \psi (4415)$, $D^*\bar{D} \to \psi (4415)$,
$D\bar{D}^* \to \psi (4415)$, $D^*\bar{D}^* \to \psi (4415)$,
$D_s^+D_s^- \to \psi (4415)$, $D_s^{*+}D_s^- \to \psi (4415)$,
$D_s^+D_s^{*-} \to \psi (4415)$, and $D_s^{*+}D_s^{*-} \to \psi (4415)$.
The $\psi (3770)$, $\psi (4040)$, $\psi (4160)$, and $\psi (4415)$ mesons
consist of a quark and an antiquark \cite{BT,GI,BGS,VFV}. 
All these reactions are governed by the strong interaction.
The reaction $K\bar{K} \to \phi$ was studied in Ref. 
\cite{LK} in a mesonic model. The twenty-one reactions that lead to  
$\psi (3770)$, $\psi (4040)$, 
$\psi (4160)$, or $\psi (4415)$ as a final state have not been studied 
theoretically. We now study $K\bar{K} \to \phi$, $\pi D \to D^*$, 
$\pi \bar{D} \to \bar{D}^*$, and the twenty-one reactions using quark degrees
of freedom. The production of $J/\psi$ is a subject intensively studied in
relativistic heavy-ion collisions. The $\psi (3770)$, $\psi (4040)$, 
$\psi (4160)$, and $\psi (4415)$ mesons
may decay into the $J/\psi$ meson. Through this decay the twenty-one reactions
add a contribution to the $J/\psi$ production in relativistic heavy-ion 
collisions. This is another reason why we study the twenty-one reactions here.

This paper is organized as follows. In Sect.~II we consider four Feynman 
diagrams
and the $S$-matrix element for 2-to-1 meson-meson scattering, derive transition
amplitudes and provide cross-section formulas. In Sect.~III we present
transition potentials corresponding to the Feynman diagrams
and calculate flavor matrix elements and spin matrix
elements. In Sect.~IV we calculate cross sections, present numerical results
and give relevant discussions. In Sect.~V we summarize the present work.
In an appendix we study decays of the $\psi (3770)$, $\psi (4040)$, 
$\psi (4160)$, and $\psi (4415)$ mesons to charmed mesons or charmed strange
mesons.

\vspace{0.5cm}
\leftline{\bf II. FORMALISM}
\vspace{0.5cm}

Lowest-order Feynman diagrams are shown in 
Fig. 1 for the reaction $A(q_1\bar{q}_1)+B(q_2\bar{q}_2) \to 
H(q_2\bar{q}_1~{\rm or}~q_1\bar{q}_2)$.
A quark in an initial meson and an antiquark in the other initial meson
annihilate into a gluon, and the gluon is then absorbed by a spectator quark
or antiquark. The four processes $q_1+\bar{q}_2+\bar{q}_1 \to \bar{q}_1$, 
$q_1+\bar{q}_2+q_2 \to q_2$, $q_2+\bar{q}_1+q_1 \to q_1$, and
$q_2+\bar{q}_1+\bar{q}_2 \to \bar{q}_2$ in Fig. 1 give rise to
the four transition potentials 
$V_{{\rm r}q_1\bar{q}_2\bar{q}_1}$, $V_{{\rm r}q_1\bar{q}_2q_2}$,
$V_{{\rm r}q_2\bar{q}_1q_1}$, and $V_{{\rm r}q_2\bar{q}_1\bar{q}_2}$, 
respectively.
Denote by $E_{\rm i}$ and $\vec{P}_{\rm i}$ ($E_{\rm f}$ and $\vec{P}_{\rm f}$)
the total energy and the total momentum of the two initial (final) mesons,
respectively; let $E_A$ ($E_B$, $E_H$) be the energy of 
meson $A$ ($B$, $H$), and
$V$ the volume where every meson wave function is normalized.
The $S$-matrix element for $A+B \to H$ is
\begin{eqnarray}
S_{\rm fi} & = & \delta_{\rm fi} - 2\pi i \delta (E_{\rm f} - E_{\rm i})
(<H \mid V_{{\rm r}q_1\bar {q}_2 \bar{q}_1}\mid A,B> + <H \mid V_{{\rm r}
q_1\bar {q}_2 q_2}\mid A,B> \nonumber\\
& &
+ <H \mid V_{{\rm r}q_2 \bar {q}_1 q_1}\mid A,B> + <H \mid V_{{\rm r}
q_2 \bar {q}_1 \bar{q}_2}\mid A,B>) \nonumber\\
& = & \delta_{\rm fi} -(2\pi)^4 i \delta (E_{\rm f} - E_{\rm i})
\delta^3 (\vec{P}_{\rm f} - \vec{P}_{\rm i})
\frac {{\cal M}_{{\rm r}q_1\bar{q}_2 \bar{q}_1}+
{\cal M}_{{\rm r}q_1\bar{q}_2 q_2}+{\cal M}_{{\rm r}q_2 \bar{q}_1 q_1}
+{\cal M}_{{\rm r}q_2 \bar{q}_1 \bar{q}_2}}
{V^{\frac {3}{2}}\sqrt{2E_A2E_B2E_H}},
\end{eqnarray}
where in the four processes mesons $A$ and $B$ go from
the state vector $\mid A,B>$ to the state vector $\mid H>$ of meson $H$, and
${\cal M}_{{\rm r}q_1\bar{q}_2 \bar{q}_1}$, 
${\cal M}_{{\rm r}q_1\bar{q}_2 q_2}$, ${\cal M}_{{\rm r}q_2 \bar{q}_1 q_1}$,
and ${\cal M}_{{\rm r}q_2 \bar{q}_1 \bar{q}_2}$ are 
the transition amplitudes given by
\begin{equation}
{\cal M}_{{\rm r}q_1\bar {q}_2 \bar{q}_1}=
\sqrt {2E_A2E_B2E_H}
\int d\vec{r}_{q_1\bar{q}_1} d\vec{r}_{q_2\bar{q}_2}
\psi_H^+ V_{{\rm r}q_1\bar{q}_2\bar{q}_1}
\psi_{AB} e^{i\vec{p}_{q_1\bar{q}_1,q_2\bar{q}_2}
\cdot\vec{r}_{q_1\bar{q}_1,q_2\bar{q}_2}},
\end{equation}
\begin{equation}
{\cal M}_{{\rm r}q_1\bar {q}_2 q_2}=
\sqrt {2E_A2E_B2E_H}
\int d\vec{r}_{q_1\bar{q}_1} d\vec{r}_{q_2\bar{q}_2}
\psi_H^+ V_{{\rm r}q_1\bar{q}_2 q_2}
\psi_{AB} e^{i\vec{p}_{q_1\bar{q}_1,q_2\bar{q}_2}
\cdot\vec{r}_{q_1\bar{q}_1,q_2\bar{q}_2}},
\end{equation}
\begin{equation}
{\cal M}_{{\rm r}q_2 \bar{q}_1 q_1}=
\sqrt {2E_A2E_B2E_H}
\int d\vec{r}_{q_1\bar{q}_1} d\vec{r}_{q_2\bar{q}_2}
\psi_H^+ V_{{\rm r}q_2\bar{q}_1 q_1}
\psi_{AB} e^{i\vec{p}_{q_1\bar{q}_1,q_2\bar{q}_2}
\cdot\vec{r}_{q_1\bar{q}_1,q_2\bar{q}_2}},
\end{equation}
\begin{equation}
{\cal M}_{{\rm r}q_2 \bar{q}_1 \bar{q}_2}=
\sqrt {2E_A2E_B2E_H}
\int d\vec{r}_{q_1\bar{q}_1} d\vec{r}_{q_2\bar{q}_2}
\psi_H^+ V_{{\rm r}q_2 \bar{q}_1 \bar{q}_2}
\psi_{AB} e^{i\vec{p}_{q_1\bar{q}_1,q_2\bar{q}_2}
\cdot\vec{r}_{q_1\bar{q}_1,q_2\bar{q}_2}},
\end{equation}
where $\vec {r}_{ab}$ is the relative coordinate of constituents $a$ and $b$;
$\vec {r}_{q_1\bar {q}_1,q_2\bar {q}_2}$ the relative coordinate 
of $q_1\bar {q}_1$ and $q_2\bar {q}_2$; 
$\vec {p}_{q_1\bar {q}_1,q_2\bar {q}_2}$ the relative momentum of 
$q_1\bar {q}_1$ and $q_2\bar {q}_2$; $\psi_H^+$ the Hermitean conjugate of 
$\psi_H$. 

The wave function of mesons $A$ and $B$ is
\begin{equation}
\psi_{AB} =\phi_{A\rm rel} \phi_{B\rm rel} \phi_{A\rm color} \phi_{B\rm color}
\chi_{S_A S_{Az}} \chi_{S_B S_{Bz}} \varphi_{AB\rm flavor},
\end{equation}
and the wave function of meson $H$ is
\begin{equation}
\psi_H =\phi_{J_HJ_{Hz}} \phi_{H\rm color} \phi_{H\rm flavor},
\end{equation}
where $S_A$ ($S_B$) is the spin of meson $A$ ($B$) with its 
magnetic projection quantum number $S_{Az}$ ($S_{Bz}$); $\phi_{A\rm rel}$ 
($\phi_{B\rm rel}$), $\phi_{A\rm color}$ ($\phi_{B\rm color}$), 
and $\chi_{S_A S_{Az}}$ ($\chi_{S_B S_{Bz}}$)
are the quark-antiquark relative-motion wave function, the color wave function,
and the spin wave function of meson $A$ ($B$), respectively;
$\phi_{H\rm flavor}$ and $\varphi_{AB\rm flavor}$ are the flavor wave 
functions of meson $H$ and of mesons $A$ and $B$, respectively;
$\phi_{H\rm color}$ is the color wave function of meson $H$; $\phi_{J_HJ_{Hz}}$
is the space-spin wave function of meson $H$ with the total angular momentum
$J_H$ and its $z$ component $J_{Hz}$.

The development in spherical harmonics of the relative-motion wave function
of mesons $A$ and $B$ (aside from a normalization constant) is given by
\begin{eqnarray}
e^{i\vec{p}_{q_1\bar{q}_1,q_2\bar{q}_2} \cdot 
\vec{r}_{q_1\bar{q}_1,q_2\bar{q}_2}} & = & 4\pi 
\sum\limits_{L_{\rm i}=0}^{\infty} 
\sum\limits_{M_{\rm i}=-L_{\rm i}}^{L_{\rm i}}
i^{L_{\rm i}} j_{L_{\rm i}} (\mid \vec{p}_{q_1\bar{q}_1,q_2\bar{q}_2} \mid
r_{q_1\bar{q}_1,q_2\bar{q}_2}) 
\nonumber     \\
& & Y_{L_{\rm i}M_{\rm i}}^\ast
(\hat{p}_{q_1\bar{q}_1,q_2\bar{q}_2}) Y_{L_{\rm i}M_{\rm i}}
(\hat{r}_{q_1\bar{q}_1,q_2\bar{q}_2}),
\end{eqnarray}
where $Y_{L_{\rm i}M_{\rm i}}$ are the spherical harmonics with 
the orbital-angular-momentum quantum number $L_{\rm i}$ and the magnetic
projection quantum number $M_{\rm i}$, $j_{L_{\rm i}}$ are the spherical Bessel
functions, and $\hat{p}_{q_1\bar{q}_1,q_2\bar{q}_2}$ 
($\hat{r}_{q_1\bar{q}_1,q_2\bar{q}_2}$)
denote the polar angles of $\vec{p}_{q_1\bar{q}_1,q_2\bar{q}_2}$
($\vec{r}_{q_1\bar{q}_1,q_2\bar{q}_2}$).
Let $\chi_{SS_{z}}$ stand for the spin wave function of mesons $A$ and $B$,
which has the total spin $S$ and its $z$ component $S_z$. The 
Clebsch-Gordan coefficients $(S_{A}S_{Az}S_{B}S_{Bz}|SS_{z})$ couple 
$\chi_{SS_z}$ to $\chi_{S_AS_{Az}}\chi_{S_BS_{Bz}}$,
\begin{equation}
  \chi_{S_{A}S_{A_{z}}}\chi_{S_{B}S_{B_{z}}}=
\sum^{S_{\rm max}}_{S=S_{\rm min}}\sum^{S}_{S_{z}=-S}
(S_{A}S_{Az}S_{B}S_{Bz}|SS_{z})\chi_{SS_{z}},
\end{equation}
where $S_{\rm min}=\mid S_A-S_B \mid$ and $S_{\rm max}=S_A+S_B$.
$Y_{L_{\rm i}M_{\rm i}}$ and $\chi_{SS_{z}}$ are coupled to the wave function
$\phi^{\rm in}_{JJ_z}$ which has the total angular momentum $J$ of mesons
$A$ and $B$ and its $z$ component $J_z$,
\begin{equation}
  Y_{L_{\rm{i}}M_{\rm{i}}}\chi_{SS_{z}}=
\sum^{J_{\rm max}}_{J=J_{\rm min}}\sum^J_{J_z=-J}
(L_{\rm{i}}M_{\rm{i}}SS_{z}|JJ_{z})\phi^{\rm{in}}_{JJ_{z}},
\end{equation}
where $J_{\rm min}=\mid L_{\rm i}-S \mid$, $J_{\rm max}=L_{\rm i}+S$, and
$(L_{\rm{i}}M_{\rm{i}}SS_{z}|JJ_{z})$ are the Clebsch-Gordan coefficients. It
follows from Eqs. (8)-(10) that the transition amplitude given in Eq. (2)
becomes
\begin{eqnarray}
  {\cal{M}}_{rq_{1}\bar{q}_{2}\bar{q}_1}&=&\sqrt{2E_{A}2E_{B}2E_{H}}4\pi
\sum^{\infty}_{L_{\rm{i}}=0}
  \sum^{L_{\rm{i}}}_{M_{\rm{i}}=-L_{\rm{i}}}i^{L_{\rm{i}}}
Y^*_{L_{\rm{i}}M_{\rm{i}}}(\hat{p}_{q_{1}\bar{q}_{1},q_{2}\bar{q}_{2}})
\phi^+_{H\rm{color}}\phi^+_{H\rm{flavor}}\nonumber\\
  &&
  \int{d^{3}r_{q_{1}\bar{q}_{1}}d^{3}r_{q_{2}\bar{q}_{2}}}
\phi^+_{J_{H}J_{Hz}}V_{rq_{1}\bar{q}_{2}\bar{q}_1}
\sum^{S_{\rm max}}_{S=S_{\rm min}}\sum^{S}_{S_{z}=-S}
(S_{A}S_{Az}S_{B}S_{Bz}|SS_{z})\nonumber\\
&&
\sum^{J_{\rm max}}_{J=J_{\rm min}}\sum^J_{J_z=-J}
(L_{\rm{i}}M_{\rm{i}}SS_{z}|JJ_{z})
\phi^{\rm{in}}_{JJ_{z}}j_{L_{\rm{i}}}(|\vec{p}_{q_{1}\bar{q}_{1},
q_{2}\bar{q}_{2}}|r_{q_{1}\bar{q}_{1},q_{2}\bar{q}_{2}})\nonumber\\
&&
\phi_{A\rm{rel}}\phi_{B\rm{rel}}\phi_{A\rm{color}}\phi_{B\rm{color}}
\varphi_{AB\rm{flavor}},
\end{eqnarray}
Denote by $L_H$ and $S_H$ the orbital angular momentum and the spin of meson
$H$, respectively, and by $M_H$ and $S_{Hz}$ the magnetic projection
quantum number of $L_H$ and $S_H$. In Eq. (11)
$\phi_{J_{H}J_{Hz}}=R_{L_H}(r_{q_2\bar{q}_1})$ $\sum^{L_H}_{M_H=-L_H}
\sum^{S_H}_{S_{Hz}=-S_H} (L_HM_HS_HS_{Hz} \mid J_HJ_{Hz}) Y_{L_HM_H}
\chi_{S_HS_{Hz}}$ where $R_{L_H}(r_{q_2\bar{q}_1})$ is the radial wave function
of the relative motion of $q_2$ and $\bar{q}_1$, and 
$(L_H M_H S_H S_{Hz} \mid J_H J_{Hz})$ are the Clebsch-Gordan coefficients.
Conservation of total angular momentum implies that $J$ equals $J_H$ and $J_z$
equals $J_{Hz}$. This leads to
\begin{eqnarray}
{\cal{M}}_{rq_{1}\bar{q}_{2}\bar{q}_1}&=&\sqrt{2E_{A}2E_{B}2E_{H}}4
\pi\sum^\infty_{L_{\rm{i}}=0}
\sum^{L_{\rm{i}}}_{M_{\rm{i}}=-L_{\rm{i}}}i^{L_{\rm{i}}}
Y^*_{L_{\rm{i}}M_{\rm{i}}}(\hat{p}_{q_{1}\bar{q}_{1},q_{2}\bar{q}_{2}})
\phi^+_{H\rm{color}}\phi^+_{H\rm{flavor}}\nonumber\\
&&
\int{d^{3}r_{q_{1}\bar{q}_{1}}d^{3}r_{q_{2}\bar{q}_{2}}}
\phi^+_{J_HJ_{Hz}}V_{rq_{1}\bar{q}_{2}\bar{q}_1}
\sum^{S_{\rm max}}_{S=S_{\rm min}}\sum^{S}_{S_{z}=-S}
(S_{A}S_{Az}S_{B}S_{Bz}|SS_{z})\nonumber\\
&&
(L_{\rm i}M_{\rm i}SS_{z}|J_{H}J_{H_{z}})\phi^{\rm{in}}_{J_{H}J_{Hz}}
j_{L_{\rm{i}}}(|\vec{p}_{q_{1}\bar{q}_{1},q_{2}\bar{q}_{2}}
|r_{q_{1}\bar{q}_{1},q_{2}\bar{q}_{2}})\nonumber\\
&&
\phi_{A\rm{rel}}\phi_{B\rm{rel}}\phi_{A\rm{color}}
\phi_{B\rm{color}}\varphi_{AB\rm{flavor}}.
\end{eqnarray}
Using the relation
\begin{equation}
  \phi^{\rm{in}}_{J_{H}J_{Hz}}=
\sum^{L_{\rm i}}_{\bar{M}_{\rm i}=-L_{\rm i}}\sum^S_{\bar{S}_{z}=-S}
(L_{\rm{i}}\bar{M}_{\rm{i}}S\bar{S}_{z}|J_{H}J_{Hz})
  Y_{L_{\rm{i}}\bar{M}_{\rm{i}}}\chi_{S\bar{S}_{z}},
\end{equation}
where $(L_{\rm{i}}\bar{M}_{\rm{i}}S\bar{S}_{z}|J_{H}J_{Hz})$ are the
Clebsch-Gordan coefficients, we get
\begin{eqnarray}
{\cal{M}}_{rq_{1}\bar{q}_{2}\bar{q}_1}
&=&\sqrt{2E_{A}2E_{B}2E_{H}}4\pi
\sum^{S_{\rm max}}_{S=S_{\rm min}}\sum^{S}_{S_{z}=-S}
(S_{A}S_{Az}S_{B}S_{Bz}
|SS_{z})\sum^\infty_{L_{\rm{i}}=0}\sum^{L_{\rm{i}}}_{M_{\rm{i}}
=-L_{\rm{i}}}i^{L_{\rm{i}}}\nonumber\\
&&
Y^*_{L_{\rm{i}}M_{\rm{i}}}(\hat{p}_{q_{1}\bar{q}_{1},q_{2}\bar{q}_{2}})
(L_{\rm{i}}M_{\rm{i}}SS_{z}|J_{H}J_{Hz})
\sum^{L_{\rm i}}_{\bar{M}_{\rm i}=-L_{\rm i}}\sum^S_{\bar{S}_{z}=-S}
(L_{\rm{i}}\bar{M}_{\rm{i}}S\bar{S}_{z}|J_{H}J_{Hz})\nonumber\\
&&
\phi^+_{H\rm{color}}\phi^+_{H\rm{flavor}}\int{d^{3}r_{q_{1}
\bar{q}_{1}}d^{3}r_{q_{2}\bar{q}_{2}}}\phi^+_{J_HJ_{Hz}}
V_{rq_{1}\bar{q}_{2}\bar{q}_1}j_{L_{\rm{i}}}
(|\vec{p}_{q_{1}\bar{q}_{1},q_{2}\bar{q}_{2}}|r_{q_{1}\bar{q}_{1},
q_{2}\bar{q}_{2}})\nonumber\\
&&
Y_{L_{\rm{i}}\bar{M}_{\rm{i}}}(\hat{r}_{q_{1}\bar{q}_{1},q_{2}\bar{q}_{2}})
\phi_{A\rm{rel}}\phi_{B\rm{rel}}\chi_{S\bar{S}_{z}}\phi_{A\rm{color}}
\phi_{B\rm{color}}\varphi_{AB\rm{flavor}}.
\end{eqnarray}
Furthermore, we need the identity
\begin{equation}
j_l(pr)Y_{lm}(\hat{r})=\int \frac{d^3p^\prime}{(2\pi)^3}\frac{2\pi^2}
{p^2}
\delta (p-p^\prime) i^l (-1)^l Y_{lm}(\hat{p}^\prime) 
e^{i\vec{p}^{~\prime} \cdot \vec{r}} ,
\end{equation}
which is obtained with the help of 
$\int_0^\infty j_l(pr)j_l(p^\prime r)r^2dr=\frac{\pi}{2p^2}
\delta (p-p^\prime)$ \cite{AW,joachain}. Substituting Eq. (15) in Eq. (14),
we get
\begin{eqnarray}
{\cal{M}}_{rq_{1}\bar{q}_{2}\bar{q}_1}&=&\sqrt{2E_{A}2E_{B}2E_{H}}4\pi
\sum^{S_{\rm max}}_{S=S_{\rm min}}\sum^{S}_{S_{z}=-S}
(S_{A}S_{Az}S_{B}S_{Bz}|SS_{z})\sum^\infty_{L_{\rm{i}}=0}
\sum^{L_{\rm{i}}}_{M_{\rm{i}}=-L_{\rm{i}}}i^{L_{\rm{i}}}\nonumber\\
&&
Y^*_{L_{\rm{i}}M_{\rm{i}}}(\hat{p}_{q_{1}\bar{q}_{1},q_{2}\bar{q}_{2}})
(L_{\rm{i}}M_{\rm{i}}SS_{z}|J_{H}J_{Hz})
\sum^{L_{\rm i}}_{\bar{M}_{\rm i}=-L_{\rm i}}\sum^S_{\bar{S}_{z}=-S}
(L_{\rm{i}}\bar{M}_{\rm{i}}S\bar{S}_{z}|J_{H}J_{Hz})\nonumber\\
&&
\phi^+_{H\rm{color}}\phi^+_{H\rm{flavor}}\int{\frac{d^{3}p_{\rm{irm}}}
{(2\pi)^3}}\frac{2\pi^2}{\vec{p}^{~2}_{q_{1}\bar{q}_{1},q_{2}\bar{q}_{2}}}
\delta(|\vec{p}_{q_{1}\bar{q}_{1},q_{2}\bar{q}_{2}}|-|\vec{p}_{\rm{irm}}|)
i^{L_{\rm{i}}}(-1)^{L_{\rm{i}}}\nonumber\\
&&
Y_{L_{\rm{i}}\bar{M}_{\rm{i}}}(\hat{p}_{\rm{irm}})\int{d^{3}r_{q_{1}
\bar{q}_{1}}d^{3}r_{q_{2}\bar{q}_{2}}}\phi^+_{J_{H}J_{Hz}}V_{rq_{1}\bar{q}_{2}
\bar{q}_{1}}e^{i\vec{p}_{\rm{irm}}\cdot\vec{r}_{q_{1}\bar{q}_{1},q_{2}
\bar{q}_{2}}}\nonumber\\
&&
\phi_{A\rm{rel}}
\phi_{B\rm{rel}}\chi_{S\bar{S}_{z}}\phi_{A\rm{color}}\phi_{B\rm{color}}
\varphi_{AB\rm{flavor}}.
\end{eqnarray}

Let $\vec{r}_c$ and $m_c$ be the position vector and the mass of constituent 
$c$, respectively. Then $\phi_{A\rm{rel}}$, $\phi_{B\rm{rel}}$, and 
$R_{L_H}$ are functions of the relative coordinate of the quark and
the antiquark. We take the Fourier transform of 
$V_{{\rm r}q_1\bar{q}_2\bar{q}_1}$ and the mesonic quark-antiquark
relative-motion wave functions:
\begin{equation}
V_{{\rm r}q_1\bar{q}_2\bar{q}_1}(\vec{r}_{\bar{q}_1}-\vec{r}_{q_1}) =
\int \frac {d^3k}{(2\pi)^3} V_{{\rm r}q_1\bar{q}_2\bar{q}_1} (\vec {k})
e^{i\vec {k} \cdot (\vec{r}_{\bar{q}_1}-\vec{r}_{q_1})},
\end{equation}
\begin{equation}
\phi_{A\rm rel}(\vec{r}_{q_1\bar{q}_1}) =
\int \frac {d^3p_{q_1\bar{q}_1}}{(2\pi)^3} \phi_{A\rm rel}
(\vec {p}_{q_1\bar{q}_1})
e^{i\vec {p}_{q_1\bar{q}_1} \cdot \vec {r}_{q_1\bar{q}_1}},
\end{equation}
\begin{equation}
\phi_{B\rm rel}(\vec{r}_{q_2\bar{q}_2}) =
\int \frac {d^3p_{q_2\bar{q}_2}}{(2\pi)^3} \phi_{B\rm rel}
(\vec {p}_{q_2\bar{q}_2})
e^{i\vec {p}_{q_2\bar{q}_2} \cdot \vec {r}_{q_2\bar{q}_2}},
\end{equation}
\begin{equation}
\phi_{J_HJ_{Hz}}(\vec{r}_{q_2\bar{q}_1}) =
\int \frac {d^3p_{q_2\bar{q}_1}}{(2\pi)^3} \phi_{J_HJ_{Hz}}
(\vec {p}_{q_2\bar{q}_1})
e^{i\vec {p}_{q_2\bar{q}_1} \cdot \vec {r}_{q_2\bar{q}_1}},
\end{equation}
for the two upper diagrams in Fig. 1, and
\begin{equation}
\phi_{J_HJ_{Hz}}(\vec{r}_{q_1\bar{q}_2}) =
\int \frac {d^3p_{q_1\bar{q}_2}}{(2\pi)^3} \phi_{J_HJ_{Hz}}
(\vec {p}_{q_1\bar{q}_2})
e^{i\vec {p}_{q_1\bar{q}_2} \cdot \vec {r}_{q_1\bar{q}_2}},
\end{equation}
for the two lower diagrams. In Eq. (17) $\vec k$ is the gluon momentum, 
and in Eqs. (18)-(21)
$\vec{p}_{ab}$ is the relative momentum of constituents $a$ and $b$. The
spherical polar coordinates of $\vec{p}_{\rm irm}$ are expressed as 
$(\mid \vec{p}_{\rm irm} \mid, \theta_{\rm irm}, \phi_{\rm irm})$. In momentum
space the normalizations are 
\begin{displaymath}
\int \frac{d^3p_{q_1\bar{q}_1}}{(2\pi)^3}
\phi_{A\rm rel}^+(\vec{p}_{q_1\bar{q}_1})
\phi_{A\rm rel}(\vec{p}_{q_1\bar{q}_1})=1, 
\end{displaymath}
\begin{displaymath}
\int \frac{d^3p_{q_2\bar{q}_2}}{(2\pi)^3}
\phi_{B\rm rel}^+(\vec{p}_{q_2\bar{q}_2})
\phi_{B\rm rel}(\vec{p}_{q_2\bar{q}_2})=1,
\end{displaymath}
\begin{displaymath}
\int \frac{d^3p_{q_2\bar{q}_1}}{(2\pi)^3}
\phi_{J_HJ_{Hz}}^+(\vec{p}_{q_2\bar{q}_1})
\phi_{J_HJ_{Hz}}(\vec{p}_{q_2\bar{q}_1})=1,
\end{displaymath}
\begin{displaymath}
\int \frac{d^3p_{q_1\bar{q}_2}}{(2\pi)^3}
\phi_{J_HJ_{Hz}}^+(\vec{p}_{q_1\bar{q}_2})
\phi_{J_HJ_{Hz}}(\vec{p}_{q_1\bar{q}_2})=1. 
\end{displaymath}
Integration over
$\mid \vec{p}_{\rm irm} \mid$, $\vec{r}_{q_1\bar{q}_1}$, and
$\vec{r}_{q_2\bar{q}_2}$ yields
\begin{eqnarray}
{\cal{M}}_{rq_1\bar{q}_2\bar{q}_1}&=&\sqrt{2E_{A}2E_{B}2E_{H}}
\sum^{S_{\rm max}}_{S=S_{\rm min}}\sum^{S}_{S_z=-S}
(S_{A}S_{Az}S_{B}S_{Bz}|SS_{z})\sum^\infty_{L_{\rm{i}}=0}
\sum^{L_{\rm{i}}}_{M_{\rm{i}}=-L_{\rm{i}}}\nonumber\\
&&
Y^*_{L_{\rm{i}}M_{\rm{i}}}(\hat{p}_{q_{1}\bar{q}_{1},q_{2}\bar{q}_{2}})
(L_{\rm{i}}M_{\rm{i}}SS_{z}|J_{H}J_{Hz})
\sum^{L_{\rm i}}_{\bar{M}_{\rm i}=-L_{\rm i}}\sum^S_{\bar{S}_{z}=-S}
(L_{\rm{i}}\bar{M}_{\rm{i}}S\bar{S}_{z}|J_{H}J_{Hz})\nonumber\\
&&
\phi^{+}_{H\rm{color}}\phi^{+}_{H\rm{flavor}}\int{d\theta_{\rm{irm}}
d\phi_{\rm{irm}}\sin\theta_{\rm{irm}}}Y_{L_{\rm{i}}\bar{M}_{\rm{i}}}
(\hat{p}_{\rm{irm}})\nonumber\\
&&
\int{\frac{d^{3}p_{q_{1}\bar{q}_{1}}}{(2\pi)^3}}
\int{\frac{d^{3}p_{q_{2}\bar{q}_{2}}}{(2\pi)^3}}\phi^+_{J_{H}J_{Hz}}
(\vec{p}_{q_{2}\bar{q}_{2}}-\frac{m_{q_{2}}}{m_{q_2}+m_{\bar{q}_{2}}}
\vec{p}_{\rm{irm}})V_{rq_1\bar{q}_2\bar{q}_1}\nonumber\\
&&
[\vec{p}_{q_{1}\bar{q}_{1}}-\vec{p}_{q_{2}\bar{q}_{2}}+
(\frac{m_{q_{2}}}{m_{q_{2}}+m_{\bar{q}_{2}}}-\frac{m_{\bar{q}_{1}}}
{m_{q_{1}}+m_{\bar{q}_{1}}})\vec{p}_{\rm{irm}}]\nonumber\\
&&
\phi_{A\rm{rel}}(\vec{p}_{q_{1}\bar{q}_{1}})\phi_{B\rm{rel}}
(\vec{p}_{q_{2}\bar{q}_{2}})\chi_{S\bar{S}_{z}}
\phi_{A\rm{color}}\phi_{B\rm{color}}\varphi_{AB\rm{flavor}},
\end{eqnarray}
in which $\mid \vec{p}_{\rm irm} \mid = \mid 
\vec{p}_{q_1\bar{q}_1,q_2\bar{q}_2} \mid$. So far, we have obtained a new
expression of the transition amplitude from Eq. (2).

Making use of the Fourier transform of $V_{{\rm r}q_1\bar{q}_2q_2}$, 
$V_{{\rm r}q_2\bar{q}_1q_1}$, and $V_{{\rm r}q_2\bar{q}_1\bar{q}_2}$,
\begin{equation}
V_{{\rm r}q_1\bar{q}_2q_2}(\vec{r}_{q_2}-\vec{r}_{\bar{q}_2}) =
\int \frac {d^3k}{(2\pi)^3} V_{{\rm r}q_1\bar{q}_2q_2} (\vec {k})
e^{i\vec {k} \cdot (\vec{r}_{q_2}-\vec{r}_{\bar{q}_2})},
\end{equation}
\begin{equation}
V_{{\rm r}q_2\bar{q}_1q_1}(\vec{r}_{q_1}-\vec{r}_{\bar{q}_1}) =
\int \frac {d^3k}{(2\pi)^3} V_{{\rm r}q_2\bar{q}_1q_1} (\vec {k})
e^{i\vec {k} \cdot (\vec{r}_{q_1}-\vec{r}_{\bar{q}_1})},
\end{equation}
\begin{equation}
V_{{\rm r}q_2\bar{q}_1\bar{q}_2}(\vec{r}_{\bar{q}_2}-\vec{r}_{q_2}) =
\int \frac {d^3k}{(2\pi)^3} V_{{\rm r}q_2\bar{q}_1\bar{q}_2} (\vec {k})
e^{i\vec {k} \cdot (\vec{r}_{\bar{q}_2}-\vec{r}_{q_2})},
\end{equation}
from Eqs. (3)-(5) we obtain
\begin{eqnarray}
{\cal{M}}_{rq_1\bar{q}_2q_2}&=&\sqrt{2E_{A}2E_{B}2E_{H}}
\sum^{S_{\rm max}}_{S=S_{\rm min}}\sum^{S}_{S_z=-S}
(S_{A}S_{Az}S_{B}S_{Bz}|SS_{z})\sum^\infty_{L_{\rm{i}}=0}
\sum^{L_{\rm{i}}}_{M_{\rm{i}}=-L_{\rm{i}}}\nonumber\\
&&
Y^*_{L_{\rm{i}}M_{\rm{i}}}(\hat{p}_{q_{1}\bar{q}_{1},q_{2}\bar{q}_{2}})
(L_{\rm{i}}M_{\rm{i}}SS_{z}|J_{H}J_{Hz})
\sum^{L_{\rm i}}_{\bar{M}_{\rm i}=-L_{\rm i}}\sum^S_{\bar{S}_{z}=-S}
(L_{\rm{i}}\bar{M}_{\rm{i}}S\bar{S}_{z}|J_{H}J_{Hz})\nonumber\\
&&
\phi^{+}_{H\rm{color}}\phi^{+}_{H\rm{flavor}}\int{d\theta_{\rm{irm}}
d\phi_{\rm{irm}}\sin\theta_{\rm{irm}}}Y_{L_{\rm{i}}\bar{M}_{\rm{i}}}
(\hat{p}_{\rm{irm}})\nonumber\\
&&
\int{\frac{d^{3}p_{q_{1}\bar{q}_{1}}}{(2\pi)^3}}
\int{\frac{d^{3}p_{q_{2}\bar{q}_{2}}}{(2\pi)^3}}\phi^+_{J_{H}J_{Hz}}
(\vec{p}_{q_1\bar{q}_1}-\frac{m_{\bar{q}_1}}{m_{q_1}+m_{\bar{q}_1}}
\vec{p}_{\rm{irm}})V_{rq_1\bar{q}_2q_2}\nonumber\\
&&
[\vec{p}_{q_{1}\bar{q}_{1}}-\vec{p}_{q_{2}\bar{q}_{2}}+
(\frac{m_{q_{2}}}{m_{q_{2}}+m_{\bar{q}_{2}}}-\frac{m_{\bar{q}_{1}}}
{m_{q_{1}}+m_{\bar{q}_{1}}})\vec{p}_{\rm{irm}}]\nonumber\\
&&
\phi_{A\rm{rel}}(\vec{p}_{q_{1}\bar{q}_{1}})\phi_{B\rm{rel}}
(\vec{p}_{q_{2}\bar{q}_{2}})\chi_{S\bar{S}_{z}}
\phi_{A\rm{color}}\phi_{B\rm{color}}\varphi_{AB\rm{flavor}},
\end{eqnarray}
\begin{eqnarray}
{\cal{M}}_{rq_2\bar{q}_1q_1}&=&\sqrt{2E_{A}2E_{B}2E_{H}}
\sum^{S_{\rm max}}_{S=S_{\rm min}}\sum^{S}_{S_z=-S}
(S_{A}S_{Az}S_{B}S_{Bz}|SS_{z})\sum^\infty_{L_{\rm{i}}=0}
\sum^{L_{\rm{i}}}_{M_{\rm{i}}=-L_{\rm{i}}}\nonumber\\
&&
Y^*_{L_{\rm{i}}M_{\rm{i}}}(\hat{p}_{q_{1}\bar{q}_{1},q_{2}\bar{q}_{2}})
(L_{\rm{i}}M_{\rm{i}}SS_{z}|J_{H}J_{Hz})
\sum^{L_{\rm i}}_{\bar{M}_{\rm i}=-L_{\rm i}}\sum^S_{\bar{S}_{z}=-S}
(L_{\rm{i}}\bar{M}_{\rm{i}}S\bar{S}_{z}|J_{H}J_{Hz})\nonumber\\
&&
\phi^{+}_{H\rm{color}}\phi^{+}_{H\rm{flavor}}\int{d\theta_{\rm{irm}}
d\phi_{\rm{irm}}\sin\theta_{\rm{irm}}}Y_{L_{\rm{i}}\bar{M}_{\rm{i}}}
(\hat{p}_{\rm{irm}})\nonumber\\
&&
\int{\frac{d^{3}p_{q_{1}\bar{q}_{1}}}{(2\pi)^3}}
\int{\frac{d^{3}p_{q_{2}\bar{q}_{2}}}{(2\pi)^3}}\phi^+_{J_{H}J_{Hz}}
(\vec{p}_{q_{2}\bar{q}_{2}}+\frac{m_{\bar{q}_2}}{m_{q_2}+m_{\bar{q}_2}}
\vec{p}_{\rm{irm}})V_{rq_2\bar{q}_1q_1}\nonumber\\
&&
[-\vec{p}_{q_{1}\bar{q}_1}+\vec{p}_{q_{2}\bar{q}_2}+
(\frac{m_{\bar{q}_2}}{m_{q_2}+m_{\bar{q}_2}}-\frac{m_{q_1}}
{m_{q_1}+m_{\bar{q}_1}})\vec{p}_{\rm{irm}}]\nonumber\\
&&
\phi_{A\rm{rel}}(\vec{p}_{q_{1}\bar{q}_{1}})\phi_{B\rm{rel}}
(\vec{p}_{q_{2}\bar{q}_{2}})\chi_{S\bar{S}_{z}}
\phi_{A\rm{color}}\phi_{B\rm{color}}\varphi_{AB\rm{flavor}},
\end{eqnarray}
\begin{eqnarray}
{\cal{M}}_{rq_2\bar{q}_1\bar{q}_2}&=&\sqrt{2E_{A}2E_{B}2E_{H}}
\sum^{S_{\rm max}}_{S=S_{\rm min}}\sum^{S}_{S_z=-S}
(S_{A}S_{Az}S_{B}S_{Bz}|SS_{z})\sum^\infty_{L_{\rm{i}}=0}
\sum^{L_{\rm{i}}}_{M_{\rm{i}}=-L_{\rm{i}}}\nonumber\\
&&
Y^*_{L_{\rm{i}}M_{\rm{i}}}(\hat{p}_{q_{1}\bar{q}_{1},q_{2}\bar{q}_{2}})
(L_{\rm{i}}M_{\rm{i}}SS_{z}|J_{H}J_{Hz})
\sum^{L_{\rm i}}_{\bar{M}_{\rm i}=-L_{\rm i}}\sum^S_{\bar{S}_{z}=-S}
(L_{\rm{i}}\bar{M}_{\rm{i}}S\bar{S}_{z}|J_{H}J_{Hz})\nonumber\\
&&
\phi^{+}_{H\rm{color}}\phi^{+}_{H\rm{flavor}}\int{d\theta_{\rm{irm}}
d\phi_{\rm{irm}}\sin\theta_{\rm{irm}}}Y_{L_{\rm{i}}\bar{M}_{\rm{i}}}
(\hat{p}_{\rm{irm}})\nonumber\\
&&
\int{\frac{d^{3}p_{q_{1}\bar{q}_{1}}}{(2\pi)^3}}
\int{\frac{d^{3}p_{q_{2}\bar{q}_{2}}}{(2\pi)^3}}\phi^+_{J_{H}J_{Hz}}
(\vec{p}_{q_1\bar{q}_1}+\frac{m_{q_1}}{m_{q_1}+m_{\bar{q}_1}}
\vec{p}_{\rm{irm}})V_{rq_2\bar{q}_1\bar{q}_2}\nonumber\\
&&
[-\vec{p}_{q_{1}\bar{q}_1}+\vec{p}_{q_{2}\bar{q}_2}+
(\frac{m_{\bar{q}_2}}{m_{q_2}+m_{\bar{q}_2}}-\frac{m_{q_1}}
{m_{q_1}+m_{\bar{q}_1}})\vec{p}_{\rm{irm}}]\nonumber\\
&&
\phi_{A\rm{rel}}(\vec{p}_{q_{1}\bar{q}_{1}})\phi_{B\rm{rel}}
(\vec{p}_{q_{2}\bar{q}_{2}})\chi_{S\bar{S}_{z}}
\phi_{A\rm{color}}\phi_{B\rm{color}}\varphi_{AB\rm{flavor}}.
\end{eqnarray}

With these transition amplitudes the unpolarized cross section for $A+B \to H$
is
\begin{eqnarray}
\sigma^{\rm unpol} & = &
\frac {\pi\delta(E_{\rm f}-E_{\rm i})}
{4\sqrt {(P_A \cdot P_B)^2 - m_A^2m_B^2}E_H}\frac {1}{(2J_A+1)(2J_B+1)}
      \nonumber  \\
& &
\sum\limits_{J_{Az}J_{Bz}J_{Hz}}
\mid {\cal M}_{{\rm r}q_1\bar{q}_2 \bar{q}_1}+{\cal M}_{{\rm r}q_1\bar{q}_2q_2}
+{\cal M}_{{\rm r}q_2\bar{q}_1 q_1}
+{{\cal M}_{{\rm r}q_2\bar{q}_1 \bar{q}_2}}\mid^2 ,
\end{eqnarray}
where $P_A$, $m_A$, $J_A$, and $J_{Az}$ ($P_B$, $m_B$, $J_B$, and $J_{Bz}$) of 
meson $A$ ($B$) are the four-momentum, the mass, the total angular momentum, 
and its $z$ component, respectively. We calculate
the cross section in the center-of-mass frame of the two initial mesons, i.e.,
with meson $H$ at rest.

\vspace{0.5cm}
\leftline{\bf III. FLAVOR AND SPIN MATRIX ELEMENTS}
\vspace{0.5cm}

Let $p_c$ be the four-momentum of constituent $c$. The two upper diagrams
give $q_1(p_{q_1})+\bar{q}_1(p_{\bar{q}_1})+q_2(p_{q_2})
+\bar{q}_2(p_{\bar{q}_2}) \to \bar{q}_1(p_{\bar{q}_1}^\prime)
+q_2(p_{q_2}^\prime)$, and the two lower diagrams
$q_1(p_{q_1})+\bar{q}_1(p_{\bar{q}_1})+q_2(p_{q_2})
+\bar{q}_2(p_{\bar{q}_2}) \to q_1(p_{q_1}^\prime)
+\bar{q}_2(p_{\bar{q}_2}^\prime)$. 
The transition potentials $V_{{\rm r}q_1\bar{q}_2\bar{q}_1}$,
$V_{{\rm r}q_1\bar{q}_2q_2}$, $V_{{\rm r}q_2\bar{q}_1q_1}$, and
$V_{{\rm r}q_2\bar{q}_1\bar{q}_2}$ are expressed as 
\begin{eqnarray}
V_{{\rm r}q_1\bar{q}_2\bar{q}_1}(\vec{k}) & = & -
\frac{\vec{\lambda}(1)}{2}\cdot\frac{\vec{\lambda}(21)}{2}
\frac{g_{\rm s}^2}{k^2}
\left(\frac{\vec{\sigma}(21)\cdot\vec{k}}{2m_{q_1}} \right.
               \nonumber \\
& & \left. -\frac{\vec{\sigma}(1)\cdot\vec{p}_{\bar{q}_1} 
\vec{\sigma}(1)\cdot\vec{\sigma}(21)
+\vec{\sigma}(1)\cdot\vec{\sigma}(21)
\vec{\sigma}(1)\cdot\vec{p}_{\bar{q}_1}^{~\prime} }
{2m_{\bar{q}_1}}\right) ,
\end{eqnarray}
\begin{eqnarray}
V_{{\rm r}q_1\bar{q}_2q_2}(\vec{k}) & = &
\frac{\vec{\lambda}(2)}{2}\cdot\frac{\vec{\lambda}(21)}{2}
\frac{g_{\rm s}^2}{k^2}
\left(\frac{\vec{\sigma}(21)\cdot\vec{k}}{2m_{q_1}} \right.
               \nonumber \\
& & \left. -
\frac{\vec{\sigma}(2)\cdot\vec{\sigma}(21)\vec{\sigma}(2)\cdot\vec{p}_{q_2}+
\vec{\sigma}(2)\cdot\vec{p}_{q_2}^{~\prime}
\vec{\sigma}(2)\cdot\vec{\sigma}(21)}{2m_{q_2}}\right) ,
\end{eqnarray}
\begin{eqnarray}
V_{{\rm r}q_2\bar{q}_1q_1}(\vec{k}) & = &
\frac{\vec{\lambda}(1)}{2}\cdot\frac{\vec{\lambda}(12)}{2}
\frac{g_{\rm s}^2}{k^2}
\left(\frac{\vec{\sigma}(12)\cdot\vec{k}}{2m_{q_2}} \right.
               \nonumber \\
& & \left. -
\frac{\vec{\sigma}(1)\cdot\vec{\sigma}(12)\vec{\sigma}(1)\cdot\vec{p}_{q_1}+
\vec{\sigma}(1)\cdot\vec{p}_{q_1}^{~\prime}
\vec{\sigma}(1)\cdot\vec{\sigma}(12)}{2m_{q_1}}\right) ,
\end{eqnarray}
\begin{eqnarray}
V_{{\rm r}q_2\bar{q}_1\bar{q}_2}(\vec{k}) & = & -
\frac{\vec{\lambda}(2)}{2}\cdot\frac{\vec{\lambda}(12)}{2}
\frac{g_{\rm s}^2}{k^2}
\left(\frac{\vec{\sigma}(12)\cdot\vec{k}}{2m_{q_2}} \right.
               \nonumber \\
& & \left. -
\frac{\vec{\sigma}(2)\cdot\vec{p}_{\bar{q}_2} 
\vec{\sigma}(2)\cdot\vec{\sigma}(12)
+\vec{\sigma}(2)\cdot\vec{\sigma}(12)
\vec{\sigma}(2)\cdot\vec{p}_{\bar{q}_2}^{~\prime} }
{2m_{\bar{q}_2}}\right) ,
\end{eqnarray}
where $g_{\rm s}$ is the gauge coupling constant, $k$ the gluon 
four-momentum, $\vec \lambda$ the Gell-Mann matrices,
and $\vec \sigma$ the Pauli matrices.
In Eqs. (30) and (31), $\vec{\lambda}(21)$ 
($\vec{\sigma}(21)$) mean that they have matrix elements between the color 
(spin) wave functions of initial antiquark $\bar{q}_2$ and initial quark $q_1$.
In Eqs. (32) and (33), $\vec{\lambda}(12)$ 
($\vec{\sigma}(12)$) mean that they have matrix elements between the color 
(spin) wave functions of initial antiquark $\bar{q}_1$ and initial quark $q_2$.
In Eqs. (30) and (33), $\vec{\lambda}(1)$ and $\vec{\lambda}(2)$
($\vec{\sigma}(1)$ and $\vec{\sigma}(2)$) mean that they 
have matrix elements between the color (spin) wave functions of the initial 
antiquark and the final antiquark. 
In Eqs. (31) and (32), $\vec{\lambda}(2)$ and $\vec{\lambda}(1)$ 
($\vec{\sigma}(2)$ and $\vec{\sigma}(1)$) mean that they 
have matrix elements between the color (spin) wave functions of the final 
quark and the initial quark.

It is shown in Refs. \cite{GI,BGS,VFV} 
that $\psi (3770)$, $\psi (4040)$, $\psi (4160)$, and $\psi (4415)$ can be 
individually interpreted as the $1 ^3D_1$, $3 ^3S_1$,
$2 ^3D_1$, and $4 ^3S_1$ quark-antiquark states. We use the notation
$ K= \left( \begin{array}{c} K^+ \\ K^0 \end{array} \right) $,
$\bar{K}= \left( \begin{array}{c} \bar{K}^0 \\ K^- \end{array} \right)$,
$ K^*= \left( \begin{array}{c} K^{*+} \\ K^{*0} \end{array} \right) $,
$\bar{K}^*= \left( \begin{array}{c} \bar{K}^{*0} \\ K^{*-} \end{array} 
\right)$,
$ D= \left( \begin{array}{c} D^+ \\ D^0 \end{array} \right) $,
$\bar{D}= \left( \begin{array}{c} \bar{D}^0 \\ D^- \end{array} \right)$,
$ D^*= \left( \begin{array}{c} D^{*+} \\ D^{*0} \end{array} \right)$, and
$\bar{D}^*= \left( \begin{array}{c} \bar{D}^{*0} \\ D^{*-} \end{array}
\right)$.
Based on the formulas in Sect. II, we study the following reactions:
\begin{displaymath}
K\bar{K} \to \phi,~\pi D \to D^*,~\pi \bar{D} \to \bar{D}^*,
\end{displaymath}
\begin{displaymath}
D\bar{D} \to \psi (3770),~D\bar{D} \to \psi (4040),
\end{displaymath}
\begin{displaymath}
D^*\bar{D} \to \psi (4040),~D\bar{D}^* \to \psi (4040),
~D^*\bar{D}^* \to \psi (4040),~D_s^+D_s^- \to \psi (4040),
\end{displaymath}
\begin{displaymath}
D\bar{D} \to \psi (4160),~D^*\bar{D} \to \psi (4160),
~D\bar{D}^* \to \psi (4160),~D^*\bar{D}^* \to \psi (4160),
\end{displaymath}
\begin{displaymath}
D_s^+D_s^- \to \psi (4160),~D_s^{*+}D_s^- \to \psi (4160),
~D_s^+D_s^{*-} \to \psi (4160),
\end{displaymath}
\begin{displaymath}
D\bar{D} \to \psi (4415),~D^*\bar{D} \to \psi (4415),
~D\bar{D}^* \to \psi (4415),~D^*\bar{D}^* \to \psi (4415),
\end{displaymath}
\begin{displaymath}
D_s^+D_s^- \to \psi (4415),~D_s^{*+}D_s^- \to \psi (4415),
~D_s^+D_s^{*-} \to \psi (4415),~D_s^{*+}D_s^{*-} \to \psi (4415).
\end{displaymath}
From the Gell-Mann matrices and the Pauli matrices in the transition 
potentials, the expressions of the transition amplitudes in Eqs. (22) and
(26)-(28) involve color matrix elements, flavor matrix elements, and spin 
matrix elements for the above reactions. The color matrix elements in 
${\cal M}_{rq_1\bar{q}_2\bar{q}_1}$, ${\cal M}_{rq_1\bar{q}_2q_2}$,
${\cal M}_{rq_2\bar{q}_1q_1}$, and ${\cal M}_{rq_2\bar{q}_1\bar{q}_2}$ are
-$\frac{4}{3\sqrt 3}$, $\frac{4}{3\sqrt 3}$, $\frac{4}{3\sqrt 3}$, and
-$\frac{4}{3\sqrt 3}$, respectively.
While we calculate the flavor matrix elements, we keep the total isospin
of the two initial mesons the same as the isospin of the final meson.
The flavor wave functions of charmed mesons and charmed strange mesons are
$\mid D^+>=-\mid c\bar{d}>$, $\mid D^0>=\mid c\bar{u}>$, 
$\mid \bar{D}^0>=\mid u\bar{c}>$, $\mid D^->=\mid d\bar{c}>$, 
$\mid D^{*+}>=-\mid c\bar{d}>$, $\mid D^{*0}>=\mid c\bar{u}>$, 
$\mid \bar{D}^{*0}>=\mid u\bar{c}>$, and $\mid D^{*-}>=\mid d\bar{c}>$.
The flavor matrix element ${\cal M}_{{\rm f}K\bar{K} \to \phi}$
(${\cal M}_{{\rm f}\pi D \to D^*}$,
${\cal M}_{{\rm f}D\bar{D} \to \psi (3770)}$,
${\cal M}_{{\rm f}D_s^+D_s^- \to \psi (4040)}$)
for $K\bar{K} \to \phi$ ($\pi D \to D^*$,
$D\bar{D} \to \psi (3770)$, $D_s^+D_s^- \to \psi (4040)$) is shown in 
Table 1. The flavor matrix element for $\pi \bar{D} \to \bar{D}^*$ equals
${\cal M}_{{\rm f}\pi D \to D^*}$.
The flavor matrix elements for 
$D\bar{D} \to \psi (4040)$, $D^*\bar{D} \to \psi (4040)$,
$D\bar{D}^* \to \psi (4040)$, $D^*\bar{D}^* \to \psi (4040)$,
$D\bar{D} \to \psi (4160)$, $D^*\bar{D} \to \psi (4160)$, 
$D\bar{D}^* \to \psi (4160)$, $D^*\bar{D}^* \to \psi (4160)$,
$D\bar{D} \to \psi (4415)$, $D^*\bar{D} \to \psi (4415)$, 
$D\bar{D}^* \to \psi (4415)$, and $D^*\bar{D}^* \to \psi (4415)$ are the same 
as ${\cal M}_{{\rm f}D\bar{D} \to \psi (3770)}$. 
The flavor matrix elements for $D_s^+D_s^- \to \psi (4160)$,
$D_s^{*+}D_s^- \to \psi (4160)$, $D_s^+D_s^{*-} \to \psi (4160)$, 
$D_s^+D_s^- \to \psi (4415)$, $D_s^{*+}D_s^- \to \psi (4415)$, 
$D_s^+D_s^{*-} \to \psi (4415)$, and $D_s^{*+}D_s^{*-} \to \psi (4415)$ equal 
${\cal M}_{{\rm f}D_s^+D_s^- \to \psi (4040)}$. The flavor matrix elements
for $K\bar{K} \to \phi$, $\pi D \to D^*$, and $\pi \bar{D} \to \bar{D}^*$ are
zero for the two lower diagrams, and those for the production of 
$\psi (3770)$, $\psi (4040)$,
$\psi(4160)$, and $\psi(4415)$ are zero for the two upper diagrams. Hence,
every reaction receives contributions only from two Feynman diagrams.

Now we give the spin matrix elements.
Let $P_{rq_1\bar{q}_2\bar{q}_1i}$ with $i=0$, $\cdot \cdot \cdot$, and 15 stand
for 1, $\sigma_1(21)$, $\sigma_2(21)$, $\sigma_3(21)$, $\sigma_1(1)$, 
$\sigma_2(1)$, $\sigma_3(1)$, $\sigma_1(21) \sigma_1(1)$, 
$\sigma_1(21) \sigma_2(1)$, $\sigma_1(21) \sigma_3(1)$, 
$\sigma_2(21) \sigma_1(1)$, $\sigma_2(21) \sigma_2(1)$, 
$\sigma_2(21) \sigma_3(1)$, $\sigma_3(21) \sigma_1(1)$, 
$\sigma_3(21) \sigma_2(1)$, and $\sigma_3(21) \sigma_3(1)$, respectively. 
Let $P_{rq_1\bar{q}_2q_2i}$ with $i=0$, $\cdot \cdot \cdot$, and 15 correspond 
to 1, $\sigma_1(21)$, $\sigma_2(21)$, $\sigma_3(21)$, $\sigma_1(2)$, 
$\sigma_2(2)$, $\sigma_3(2)$, $\sigma_1(21) \sigma_1(2)$, 
$\sigma_1(21) \sigma_2(2)$, $\sigma_1(21) \sigma_3(2)$, 
$\sigma_2(21) \sigma_1(2)$, $\sigma_2(21) \sigma_2(2)$, 
$\sigma_2(21) \sigma_3(2)$, $\sigma_3(21) \sigma_1(2)$, 
$\sigma_3(21) \sigma_2(2)$, and $\sigma_3(21) \sigma_3(2)$, respectively. 
Let $P_{rq_2\bar{q}_1q_1i}$ with $i=0$, $\cdot \cdot \cdot$, and 15 
represent 1, $\sigma_1(12)$, $\sigma_2(12)$, $\sigma_3(12)$, $\sigma_1(1)$, 
$\sigma_2(1)$, $\sigma_3(1)$, $\sigma_1(12) \sigma_1(1)$, 
$\sigma_1(12) \sigma_2(1)$, $\sigma_1(12) \sigma_3(1)$, 
$\sigma_2(12) \sigma_1(1)$, $\sigma_2(12) \sigma_2(1)$, 
$\sigma_2(12) \sigma_3(1)$, $\sigma_3(12) \sigma_1(1)$, 
$\sigma_3(12) \sigma_2(1)$, and $\sigma_3(12)$ $\sigma_3(1)$, respectively. 
Let $P_{rq_2\bar{q}_1\bar{q}_2i}$ with $i=0$, $\cdot \cdot \cdot$, and 15
denote 1, $\sigma_1(12)$, $\sigma_2(12)$, $\sigma_3(12)$, $\sigma_1(2)$, 
$\sigma_2(2)$, $\sigma_3(2)$, $\sigma_1(12) \sigma_1(2)$, 
$\sigma_1(12) \sigma_2(2)$, $\sigma_1(12) \sigma_3(2)$, 
$\sigma_2(12) \sigma_1(2)$, $\sigma_2(12) \sigma_2(2)$, 
$\sigma_2(12)$ $\sigma_3(2)$, $\sigma_3(12) \sigma_1(2)$, 
$\sigma_3(12) \sigma_2(2)$, and $\sigma_3(12) \sigma_3(2)$, respectively. 
Set $n_A$ as -1, 0, or 1, and $n_B$ as -1, 0, or 1. In order to easily tabulate
values
of the spin matrix elements, we define $\phi_{\rm iss} (S_A,S_{Az};S_B,S_{Bz}) 
\equiv \chi_{S_AS_{Az}}\chi_{S_BS_{Bz}}$ and 
$\phi_{\rm fss}(S_H,S_{Hz})\equiv \chi_{S_HS_{Hz}}$.
The spin matrix 
elements $\phi^+_{\rm fss}(S_H,S_{Hz})P_{rq_1\bar{q}_2\bar{q}_1i}
\phi_{\rm iss} (S_A,S_{Az};S_B,S_{Bz})$ are shown in Tables 2-6. Other
spin matrix elements, $\phi^+_{\rm fss} (S_H, S_{Hz}) P_{rq_1\bar{q}_2q_2i}
\phi_{\rm iss} (S_A, S_{Az}; S_B,$ $S_{Bz})$,
$\phi^+_{\rm fss} (S_H, S_{Hz})$ $P_{rq_2\bar{q}_1q_1i}$
$\phi_{\rm iss} (S_A, S_{Az}; S_B, S_{Bz})$, and
$\phi^+_{\rm fss} (S_H, S_{Hz})$ $P_{rq_2\bar{q}_1\bar{q}_2i}$
$\phi_{\rm iss} (S_A, S_{Az};$ $S_B, S_{Bz})$, are related to 
$\phi^+_{\rm fss} (S_H, S_{Hz}) P_{rq_1\bar{q}_2\bar{q}_1i}
\phi_{\rm iss} (S_A, S_{Az}; S_B, S_{Bz})$ by the following equations:
\begin{eqnarray}
& & \phi^+_{\rm fss}(S_H=1,S_{Hz})P_{rq_1\bar{q}_2q_2i}\phi_{\rm iss} 
(S_A=1,S_{Az}=n_A; S_B=0,S_{Bz}=0)
\nonumber  \\
& = & \phi^+_{\rm fss}(S_H=1,S_{Hz})P_{rq_1\bar{q}_2\bar{q}_1i}\phi_{\rm iss} 
(S_A=0,S_{Az}=0; S_B=1,S_{Bz}=n_A) ,
\end{eqnarray}
with $i=2$, 5, 8, 10, 12, and 14;
\begin{eqnarray}
& & \phi^+_{\rm fss}(S_H=1,S_{Hz})P_{rq_1\bar{q}_2q_2i}\phi_{\rm iss} 
(S_A=1,S_{Az}=n_A; S_B=0,S_{Bz}=0)
\nonumber  \\
& = & -\phi^+_{\rm fss}(S_H=1,S_{Hz})P_{rq_1\bar{q}_2\bar{q}_1i}\phi_{\rm iss} 
(S_A=0,S_{Az}=0; S_B=1,S_{Bz}=n_A) ,
\end{eqnarray}
with $i=0$, 1, 3, 4, 6, 7, 9, 11, 13, and 15;
\begin{eqnarray}
& & \phi^+_{\rm fss}(S_H=1,S_{Hz})P_{rq_1\bar{q}_2q_2i}\phi_{\rm iss} 
(S_A=0,S_{Az}=0; S_B=1,S_{Bz}=n_B)
\nonumber  \\
& = & \phi^+_{\rm fss}(S_H=1,S_{Hz})P_{rq_1\bar{q}_2\bar{q}_1i}\phi_{\rm iss} 
(S_A=1,S_{Az}=n_B; S_B=0,S_{Bz}=0) ,
\end{eqnarray}
with $i=2$, 5, 8, 10, 12, and 14;
\begin{eqnarray}
& & \phi^+_{\rm fss}(S_H=1,S_{Hz})P_{rq_1\bar{q}_2q_2i}\phi_{\rm iss} 
(S_A=0,S_{Az}=0; S_B=1,S_{Bz}=n_B)
\nonumber  \\
& = & -\phi^+_{\rm fss}(S_H=1,S_{Hz})P_{rq_1\bar{q}_2\bar{q}_1i}\phi_{\rm iss} 
(S_A=1,S_{Az}=n_B; S_B=0,S_{Bz}=0) ,
\end{eqnarray}
with $i=0$, 1, 3, 4, 6, 7, 9, 11, 13, and 15;
\begin{eqnarray}
& & \phi^+_{\rm fss}(S_H=1,S_{Hz})P_{rq_1\bar{q}_2q_2i}\phi_{\rm iss} 
(S_A=1,S_{Az}=n_A; S_B=1,S_{Bz}=n_B)
\nonumber  \\
& = & \phi^+_{\rm fss}(S_H=1,S_{Hz})P_{rq_1\bar{q}_2\bar{q}_1i}\phi_{\rm iss} 
(S_A=1,S_{Az}=n_B; S_B=1,S_{Bz}=n_A) ,
\end{eqnarray}
with $i=0$, 1, 3, 4, 6, 7, 9, 11, 13, and 15;
\begin{eqnarray}
& & \phi^+_{\rm fss}(S_H=1,S_{Hz})P_{rq_1\bar{q}_2q_2i}\phi_{\rm iss} 
(S_A=1,S_{Az}=n_A; S_B=1,S_{Bz}=n_B)
\nonumber  \\
& = & -\phi^+_{\rm fss}(S_H=1,S_{Hz})P_{rq_1\bar{q}_2\bar{q}_1i}\phi_{\rm iss} 
(S_A=1,S_{Az}=n_B; S_B=1,S_{Bz}=n_A) ,
\end{eqnarray}
with $i=2$, 5, 8, 10, 12, and 14;
\begin{eqnarray}
& & \phi^+_{\rm fss}(S_H=1,S_{Hz})P_{rq_2\bar{q}_1q_1i}\phi_{\rm iss} 
(S_A=1,S_{Az}=n_A; S_B=0,S_{Bz}=0)
\nonumber  \\
& = & \phi^+_{\rm fss}(S_H=1,S_{Hz})P_{rq_1\bar{q}_2\bar{q}_1i}\phi_{\rm iss} 
(S_A=1,S_{Az}=n_A; S_B=0,S_{Bz}=0) ,
\end{eqnarray}
with $i=2$, 5, 8, 10, 12, and 14;
\begin{eqnarray}
& & \phi^+_{\rm fss}(S_H=1,S_{Hz})P_{rq_2\bar{q}_1q_1i}\phi_{\rm iss} 
(S_A=1,S_{Az}=n_A; S_B=0,S_{Bz}=0)
\nonumber  \\
& = & -\phi^+_{\rm fss}(S_H=1,S_{Hz})P_{rq_1\bar{q}_2\bar{q}_1i}\phi_{\rm iss} 
(S_A=1,S_{Az}=n_A; S_B=0,S_{Bz}=0) ,
\end{eqnarray}
with $i=0$, 1, 3, 4, 6, 7, 9, 11, 13, and 15;
\begin{eqnarray}
& & \phi^+_{\rm fss}(S_H=1,S_{Hz})P_{rq_2\bar{q}_1q_1i}\phi_{\rm iss} 
(S_A=0,S_{Az}=0; S_B=1,S_{Bz}=n_B)
\nonumber  \\
& = & \phi^+_{\rm fss}(S_H=1,S_{Hz})P_{rq_1\bar{q}_2\bar{q}_1i}\phi_{\rm iss} 
(S_A=0,S_{Az}=0; S_B=1,S_{Bz}=n_B) ,
\end{eqnarray}
with $i=2$, 5, 8, 10, 12, and 14;
\begin{eqnarray}
& & \phi^+_{\rm fss}(S_H=1,S_{Hz})P_{rq_2\bar{q}_1q_1i}\phi_{\rm iss} 
(S_A=0,S_{Az}=0; S_B=1,S_{Bz}=n_B)
\nonumber  \\
& = & -\phi^+_{\rm fss}(S_H=1,S_{Hz})P_{rq_1\bar{q}_2\bar{q}_1i}\phi_{\rm iss} 
(S_A=0,S_{Az}=0; S_B=1,S_{Bz}=n_B) ,
\end{eqnarray}
with $i=0$, 1, 3, 4, 6, 7, 9, 11, 13, and 15;
\begin{eqnarray}
& & \phi^+_{\rm fss}(S_H=1,S_{Hz})P_{rq_2\bar{q}_1q_1i}\phi_{\rm iss} 
(S_A=1,S_{Az}=n_A; S_B=1,S_{Bz}=n_B)
\nonumber  \\
& = & \phi^+_{\rm fss}(S_H=1,S_{Hz})P_{rq_1\bar{q}_2\bar{q}_1i}\phi_{\rm iss} 
(S_A=1,S_{Az}=n_A; S_B=1,S_{Bz}=n_B) ,
\end{eqnarray}
with $i=0$, 1, 3, 4, 6, 7, 9, 11, 13, and 15;
\begin{eqnarray}
& & \phi^+_{\rm fss}(S_H=1,S_{Hz})P_{rq_2\bar{q}_1q_1i}\phi_{\rm iss} 
(S_A=1,S_{Az}=n_A; S_B=1,S_{Bz}=n_B)
\nonumber  \\
& = & -\phi^+_{\rm fss}(S_H=1,S_{Hz})P_{rq_1\bar{q}_2\bar{q}_1i}\phi_{\rm iss} 
(S_A=1,S_{Az}=n_A; S_B=1,S_{Bz}=n_B) ,
\end{eqnarray}
with $i=2$, 5, 8, 10, 12, and 14;
\begin{eqnarray}
& & \phi^+_{\rm fss}(S_H=1,S_{Hz})P_{rq_2\bar{q}_1\bar{q}_2i}\phi_{\rm iss} 
(S_A=1,S_{Az}=n_A; S_B=0,S_{Bz}=0)
\nonumber  \\
& = & \phi^+_{\rm fss}(S_H=1,S_{Hz})P_{rq_1\bar{q}_2\bar{q}_1i}\phi_{\rm iss} 
(S_A=0,S_{Az}=0; S_B=1,S_{Bz}=n_A) ,
\end{eqnarray}
with $i=0$, $\cdot \cdot \cdot$, and 15;
\begin{eqnarray}
& & \phi^+_{\rm fss}(S_H=1,S_{Hz})P_{rq_2\bar{q}_1\bar{q}_2i}\phi_{\rm iss} 
(S_A=0,S_{Az}=0; S_B=1,S_{Bz}=n_B)
\nonumber  \\
& = & \phi^+_{\rm fss}(S_H=1,S_{Hz})P_{rq_1\bar{q}_2\bar{q}_1i}\phi_{\rm iss} 
(S_A=1,S_{Az}=n_B; S_B=0,S_{Bz}=0) ,
\end{eqnarray}
with $i=0$, $\cdot \cdot \cdot$, and 15;
\begin{eqnarray}
& & \phi^+_{\rm fss}(S_H=1,S_{Hz})P_{rq_2\bar{q}_1\bar{q}_2i}\phi_{\rm iss} 
(S_A=1,S_{Az}=n_A; S_B=1,S_{Bz}=n_B)
\nonumber  \\
& = & \phi^+_{\rm fss}(S_H=1,S_{Hz})P_{rq_1\bar{q}_2\bar{q}_1i}\phi_{\rm iss} 
(S_A=1,S_{Az}=n_B; S_B=1,S_{Bz}=n_A) ,
\end{eqnarray}
with $i=0$, $\cdot \cdot \cdot$, and 15.

\vspace{0.5cm}
\leftline{\bf IV. NUMERICAL CROSS SECTIONS AND DISCUSSIONS }
\vspace{0.5cm}

The mesonic quark-antiquark relative-motion wave functions $\phi_{A\rm rel}$,
$\phi_{B\rm rel}$, and $\phi_{J_HJ_{Hz}}$ in Eqs. (6) and (7)  
are solutions of the Schr\"odinger equation with the
potential between constituents $a$ and $b$ in coordinate space \cite{JSX},
\begin{eqnarray}
V_{ab}(\vec{r}_{ab}) & = &
- \frac {\vec{\lambda}_a}{2} \cdot \frac {\vec{\lambda}_b}{2}
\xi_1 \left[ 1.3- \left( \frac {T}{T_{\rm c}} \right)^4 \right] \tanh 
(\xi_2 r_{ab}) + \frac {\vec{\lambda}_a}{2} \cdot \frac {\vec{\lambda}_b}{2}
\frac {6\pi}{25} \frac {v(\lambda r_{ab})}{r_{ab}} \exp (-\xi_3 r_{ab})
\nonumber  \\
& & -\frac {\vec{\lambda}_a}{2} \cdot \frac {\vec{\lambda}_b}{2}
\frac {16\pi^2}{25}\frac{d^3}{\pi^{3/2}}\exp(-d^2r^2_{ab}) 
\frac {\vec {s}_a \cdot \vec {s}_b} {m_am_b}
+\frac {\vec{\lambda}_a}{2} \cdot \frac {\vec{\lambda}_b}{2}\frac {4\pi}{25}
\frac {1} {r_{ab}} \frac {d^2v(\lambda r_{ab})}{dr_{ab}^2} 
\frac {\vec {s}_a \cdot \vec {s}_b}{m_am_b}
\nonumber  \\
& & -\frac {\vec{\lambda}_a}{2} \cdot \frac {\vec{\lambda}_b}{2}
\frac {6\pi}{25m_am_b}\left[ v(\lambda r_{ab}) 
-r_{ab}\frac {dv(\lambda r_{ab})}{dr_{ab}} +\frac{r_{ab}^2}{3}
\frac {d^2v(\lambda r_{ab})}{dr_{ab}^2} \right]
\nonumber  \\
& & \left( \frac{3\vec {s}_a \cdot \vec{r}_{ab}\vec {s}_b \cdot \vec{r}_{ab}}
{r_{ab}^5} -\frac {\vec {s}_a \cdot \vec {s}_b}{r_{ab}^3} \right) ,
\end{eqnarray}
where $\xi_1=0.525$ GeV, $\xi_3=0.6$ GeV, $T_{\rm c}=0.175$ GeV,
$\xi_2=1.5[0.75+0.25 (T/{T_{\rm c}})^{10}]^6$ GeV, and
$\lambda=\sqrt{25/16\pi^2 \alpha'}$ with $\alpha'=1.04$ GeV$^{-2}$;
$T$ is the temperature;
$\vec {s}_a$ is the spin of constituent $a$; the quantity $d$ is given in Ref. 
\cite{JSX}; the function $v$ is given by Buchm\"uller and Tye in Ref. 
\cite{BT}. The potential is obtained from perturbative QCD \cite{BT} and
lattice QCD \cite{KLP}. The masses of the up quark, the
down quark, the strange quark, and the charm quark are 0.32 GeV, 0.32 GeV,
0.5 GeV, and 1.51 GeV, respectively. Solving the Schr\"odinger equation with
the potential at zero temperature, we obtain meson masses that are close to
the experimental masses of $\pi$, $\rho$, $K$, $K^*$, $J/\psi$, $\chi_{c}$, $
\psi'$, $\psi (3770)$, $\psi (4040)$, $\psi (4160)$, $\psi (4415)$, $D$, $D^*$,
$D_s$, and $D^*_s$ 
mesons~\cite{PDG}. Moreover, the experimental data of $S$-wave and $P$-wave
elastic phase shifts for $\pi \pi$ scattering in 
vacuum~\cite{pipiqi,pipianni} are reproduced in the Born approximation 
\cite{JSX,SXW}.

Gluon, quark, and antiquark fields in the thermal medium depend on its 
temperature. The interaction between two constituents
is influenced by gluons, quarks, and antiquarks in the thermal
medium, and thus depends on the temperature. The quark-antiquark potential is 
related to the Polyakov loop correlation function defined from the gluon
field, the quark field, and the antiquark field, and has 
been obtained in lattice gauge calculations. When the temperature is low, the
potential at large distances is modified by the medium. When the temperature is
near the critical temperature $T_{\rm c}$, the potential at intermediate 
distances is also modified. The lattice gauge calculations \cite{KLP}
only provide a numerical spin-independent and temperature-dependent
potential at intermediate and large distances. 
When the distance increases from zero, the numerical potential increases, and
obviously becomes a distance-independent value (exhibits a plateau) at
large distances at $T>0.55T_{\rm c}$. The plateau height decreases with 
increasing temperature. This means that confinement becomes weaker and weaker.
The short-distance part of the first two terms in Eq. (49) originates from
one-gluon exchange plus perturbative one- and two-loop corrections, and the
intermediate-distance and large-distance part fits well the numerical
potential. The third and fourth terms indicate the spin-spin interaction with 
relativistic effects, and the fifth term is the tensor interaction, 
which are obtained from an application of the 
Foldy-Wouthuysen canonical transformation to the gluon propagator with 
perturbative one- and two-loop corrections \cite{Xu2002}. The potential in
Eq. (49) is valid when the temperature is below the critical temperature.

In Eq. (49) $\xi_1$ is fixed, but the values of $\xi_2$ and $\xi_3$ are not 
unique; $\xi_2$ is allowed to change by at most 5\%, and $\xi_3$ by at most
10\% while the numerical potential obtained in the lattice gauge calculations
can be well fitted. However, an increase (decrease) of $\xi_2$ must be 
accompanied by a decrease (increase) of $\xi_3$. The uncertainties of
$\xi_2$ and $\xi_3$ cause a change less than 2.7\% in meson mass,
a change less than 3.6\% in cross section, and a change less than 4.3\% in
decay width.

From the transition potentials, the color matrix elements, 
the flavor matrix elements, the spin matrix
elements, and the mesonic quark-antiquark relative-motion wave functions,
we calculate the transition amplitudes. As seen in Eq. (8), the development in
spherical harmonics 
contains the summation over the orbital-angular-momentum quantum
number $L_{\rm i}$ that labels the relative motion between mesons $A$ and $B$. 
However, not all orbital-angular-momentum quantum numbers are allowed.
The orbital-angular-momentum quantum numbers are selected to satisfy
parity conservation and $J=J_H$, i.e., the total angular momentum of the two
initial mesons equals the total angular momentum of meson $H$.
The choice of $L_{\rm i}$ thus depends on the total spin $S$ of
the two initial mesons. From the transition amplitudes we get unpolarized 
cross sections at zero temperature. 
The selected orbital-angular-momentum quantum numbers 
and the cross sections are shown in Tables 7 and 8. The processes 
$D^\ast \bar{D}^\ast \to \psi (4040)$, 
$D^\ast \bar{D}^\ast \to \psi (4160)$, $D^\ast \bar{D}^\ast \to \psi (4415)$,
and $D^{\ast +}_s D^{\ast -}_s \to \psi (4415)$ allow $S=0$, $S=1$,
and $S=2$. Including contributions from $S=0$, $S=1$, and $S=2$, the cross
sections for the four reactions are 0.42 mb, 0.57 mb, 1.39 mb, and 0.11 mb, 
respectively. When $L_{\rm i}=1$ is selected, the partial wave for 
$L_{\rm i}=1$ in Eq. (8) is normalized as
\begin{displaymath}
4\pi \mid \vec{p}_{q_1\bar{q}_1,q_2\bar{q}_2} \mid
i j_1 (\mid \vec{p}_{q_1\bar{q}_1,q_2\bar{q}_2} \mid
r_{q_1\bar{q}_1,q_2\bar{q}_2}) Y_{1M_{\rm i}}
(\hat{r}_{q_1\bar{q}_1,q_2\bar{q}_2}).
\end{displaymath}
When $L_{\rm i}=1$ and $L_{\rm i}=3$ are selected together, the partial waves
for $L_{\rm i}=1$ and $L_{\rm i}=3$ are normalized as
\begin{displaymath}
4\pi \mid \vec{p}_{q_1\bar{q}_1,q_2\bar{q}_2} \mid i
j_1 (\mid \vec{p}_{q_1\bar{q}_1,q_2\bar{q}_2} \mid
r_{q_1\bar{q}_1,q_2\bar{q}_2}) \sum\limits_{M_{\rm i}=-1}^1 Y_{1M_{\rm i}}
(\hat{r}_{q_1\bar{q}_1,q_2\bar{q}_2}) \sqrt{\frac{4\pi}{10}}
Y_{1M_{\rm i}}^\ast (\hat{p}_{q_1\bar{q}_1,q_2\bar{q}_2})
\end{displaymath}
\begin{displaymath}
+4\pi \mid \vec{p}_{q_1\bar{q}_1,q_2\bar{q}_2} \mid i^3
j_3 (\mid \vec{p}_{q_1\bar{q}_1,q_2\bar{q}_2} \mid
r_{q_1\bar{q}_1,q_2\bar{q}_2}) \sum\limits_{M_{\rm i}=-3}^3 Y_{3M_{\rm i}}
(\hat{r}_{q_1\bar{q}_1,q_2\bar{q}_2}) \sqrt{\frac{4\pi}{10}}
Y_{3M_{\rm i}}^\ast (\hat{p}_{q_1\bar{q}_1,q_2\bar{q}_2}).
\end{displaymath}

In Table 7 the cross section for $K\bar{K} \to \phi$ equals 8.05 mb. The
magnitude 8.05 mb is larger than the peak cross section of
$K\bar{K} \to K^*\bar{K}^*$ for total isospin $I=0$ at zero temperature,
and is roughly 8 times the peak cross section of 
$K\bar{K} \to K^*\bar{K}^*$ for $I=1$ in Ref. \cite{SXW}. 
The case $K\bar{K} \to K^*\bar{K}^*$ may be caused by a process
where a quark in an initial meson and an antiquark in another initial meson 
annihilate into a gluon and subsequently the gluon creates another 
quark-antiquark pair. The magnitude is much larger than the peak cross
sections of $K\bar{K} \to \pi K \bar K$ for I=1 and 
$I_{\pi \bar K}^{\rm f}=3/2$ and for I=1 and $I_{\pi \bar K}^{\rm f}=1/2$
at zero temperature in Refs. \cite{LXW,XW}, 
where $I_{\pi \bar K}^{\rm f}$ is the total isospin of the final $\pi$ and 
$\bar K$ mesons. The case $K\bar{K} \to \pi K \bar K$ is
governed by a process where a gluon is emitted by a constituent quark or
antiquark in the initial mesons and subsequently the gluon creates a 
quark-antiquark pair. The magnitude is also much larger than the peak cross
section of $KK \to K^\ast K^\ast$ for $I=1$ at zero temperature in Ref.
\cite{SX1}. The case $KK \to K^\ast K^\ast$ for $I=1$ can be caused by quark 
interchange between the two colliding mesons.
The cross section for $\pi D \to D^*$ is particularly large. This 
means that the reaction easily happens. The large cross section is caused by  
the very small
difference between the $D^*$ mass and the sum of the $\pi$ and $D$ masses.

The transition potentials involve quark masses. The charm-quark mass is
larger than the strange-quark mass, and the transition potentials with
the charm quark are smaller than the ones with the strange quark. The
cross sections for $D\bar{D} \to \psi (3770)$, $D\bar{D} \to \psi (4040)$,
$D\bar{D} \to \psi (4160)$, and $D\bar{D} \to \psi (4415)$
are thus smaller than the one for $K\bar{K} \to \phi$.
Because the $D^*$ radius is larger than the $D$ radius, the cross sections for 
$D^*\bar{D} \to \psi (4040)$, $D^*\bar{D} \to \psi (4160)$, and
$D^*\bar{D} \to \psi (4415)$ are larger than those for 
$D\bar{D} \to \psi (4040)$, $D\bar{D} \to \psi (4160)$, and
$D\bar{D} \to \psi (4415)$, respectively. However, the cross sections for
$D^*\bar{D}^* \to \psi (4040)$, $D^*\bar{D}^* \to \psi (4160)$, and 
$D^*\bar{D}^* \to \psi (4415)$ are smaller than those for 
$D^*\bar{D} \to \psi (4040)$, $D^*\bar{D} \to \psi (4160)$, and
$D^*\bar{D} \to \psi (4415)$, respectively. This is caused by the two nodes
of $\psi (4040)$, the node of $\psi (4160)$, and the three nodes of 
$\psi (4415)$ in the radial part of the 
quark-antiquark relative-motion wave function.
The radial wave function on the left of a node has a sign different from the
one on the right of the node. The nodes lead to cancellation between the 
positive radial wave function and the negative radial wave function in the
integration involved in the transition amplitudes.
Since $D^+_s$ (the antiparticle of $D^-_s$) 
consists of a charm quark and a strange antiquark, the cross sections for
$D^+_s D^-_s \to \psi (4040)$, $D^+_s D^-_s \to \psi (4160)$, and 
$D^+_s D^-_s \to \psi (4415)$ are smaller than the ones for
$D\bar{D} \to \psi (4040)$, $D\bar{D} \to \psi (4160)$, and 
$D\bar{D} \to \psi (4415)$, respectively.

In the present work the $\psi (3770)$, $\psi (4040)$, $\psi (4160)$, and 
$\psi (4415)$ mesons come from
fusion of $D$, $\bar D$, $D^*$, $\bar{D}^*$, $D_s$, and $D^*_s$ mesons.
The reason why we are interested in these reactions is that $\psi (3770)$, 
$\psi (4040)$, $\psi (4160)$, and 
$\psi (4415)$ may decay into $J/\psi,$ which is an important probe of the
quark-gluon plasma produced in ultrarelativistic heavy-ion collisions. 
We do not investigate
the $\chi_{c0}(2P)$ and $\chi_{c2}(2P)$ mesons because they cannot decay into
the $J/\psi$ meson. The $\chi_{c1}(2P)$ meson may decay into the $J/\psi$
meson, but $D \bar{D} \to \chi_{c1}(2P)$ allowed 
by energy conservation does not simultaneously satisfy
the parity conservation and the conservation of the total angular momentum.
Therefore, we do not consider $D \bar{D} \to \chi_{c1}(2P)$. The production
of $\chi_{c1}(2P)$ from fusion of other charmed mesons is forbidden by 
energy conservation. Also $\pi D^{\pm}_s \to D^{*\pm}_s$ is not allowed because
of violation of isospin conservation, and $K D \to D^{*+}_s$ and
$\bar{K} \bar{D} \to D^{*-}_s$ because of violation of energy 
conservation.

As seen in Eq. (49), the potential between two constituents depends on 
temperature. The meson mass obtained from the Schr\"odinger equation with
the potential thus depends on temperature. The temperature dependence of meson 
masses is shown in Figs. 2-5. In vacuum the $\phi$ mass is larger than two
times the kaon mass, and so the reaction $K\bar{K} \to \phi$ takes place.
Since the $\phi$ mass in Fig. 2 decreases faster than the kaon mass with 
increasing temperature, the $\phi$ mass turns smaller than two times the kaon 
mass. The reaction thus no longer occurs in the temperature region
$0.6T_{\rm c}\leq T <T_{\rm c}$. In Fig. 3 the $D^*$ mass decreases
faster than the pion and $D$ masses, and the $D^*$ mass is smaller than the sum
of the pion mass and the $D$ mass. The processes 
$\pi D \to D^*$ and $\pi \bar{D} \to \bar{D}^*$ also do not occur for
$0.6T_{\rm c}\leq T <T_{\rm c}$. The mesons 
$\psi (4040)$, $\psi (4160)$, and $\psi (4415)$ are dissolved in 
hadronic matter when the temperature is larger than $0.97T_{\rm c}$,
$0.95T_{\rm c}$, and $0.87T_{\rm c}$, respectively \cite{JXW}. 
Their masses are thus plotted only for $0.6T_{\rm c}\leq T < 0.97T_{\rm c}$, 
for $0.6T_{\rm c}\leq T < 0.95T_{\rm c}$,
and for $0.6T_{\rm c}\leq T < 0.87T_{\rm c}$ 
in Figs. 4 and 5, and are smaller than the sum of the masses of the two initial
mesons that yield them. Therefore, in hadronic matter where the temperature is
constrained by $0.6T_{\rm c}\leq T < T_{\rm c}$,
we cannot see the production of $\psi (4040)$, $\psi (4160)$, and $\psi (4415)$
from the fusion of two charmed mesons and of two charmed
strange mesons. Here $T_{\rm c}$ is the critical temperature at which the phase
transition between the quark-gluon plasma and hadronic matter takes place.
Since $\psi (4040)$, $\psi (4160)$, and $\psi (4415)$ are dissolved in hadronic
matter when the temperature is larger than $0.97T_{\rm c}$, $0.95T_{\rm c}$, 
and $0.87T_{\rm c}$, 
respectively, they cannot be produced in the phase transition, 
but they can be produced in the following reactions in hadronic matter:
\begin{displaymath}
D \bar{D} \to \rho R;
~D \bar{D}^* \to \pi R,~\rho R, \eta R;
~D^* \bar{D} \to \pi R, \rho R, \eta R;
~D^* \bar{D}^* \to \pi R, \rho R, \eta R;
\end{displaymath}
\begin{displaymath}
D_s^+ \bar{D} \to K^\ast R;
~D_s^- D \to \bar{K}^\ast R;
~D_s^+ \bar{D}^* \to K R, K^\ast R;
~D_s^- D^* \to \bar{K} R, \bar{K}^\ast R;
\end{displaymath}
\begin{displaymath}
D_s^{*+} \bar{D} \to K R, K^\ast R;
~D_s^{*-} D \to \bar{K} R, \bar{K}^\ast R;
~D_s^{*+} \bar{D}^* \to K R, K^\ast R;
~D_s^{*-} D^* \to \bar{K} R, \bar{K}^\ast R;
\end{displaymath}
\begin{displaymath}
D_s^+ D_s^- \to \phi R;
~D_s^+ D_s^{*-} \to \eta R, \phi R;
~D_s^{*+} D_s^- \to \eta R, \phi R;
~D_s^{*+} D_s^{*-} \to \eta R, \phi R.
\end{displaymath}
In these reactions $R$ stands for $\psi (4040)$, $\psi (4160)$, or 
$\psi (4415)$. Hadronic matter undergoes expansion, 
and its temperature decreases until kinetic freeze-out occurs. Every kind 
of hadron in hadronic matter satisfies a momentum distribution function. 
The temperature, the expansion, and the momentum distribution 
functions are needed for a full description of hadronic matter. 
The $\psi (4040)$, $\psi (4160)$, and 
$\psi (4415)$ mesons produced from hadronic matter depend on the
temperature, the expansion, and the distribution functions. 
Inversely, from the production we know the temperature, the expansion, and 
the distribution functions.
Therefore, $\psi (4040)$, $\psi (4160)$, and $\psi (4415)$ may provide us with 
information on hadronic matter, 
and are a probe of hadronic matter that results from the quark-gluon plasma.

No experimental data on $K\bar{K} \to \phi$ are available.
Our cross section for $K\bar{K} \to \phi$ is about 8 mb, which is smaller than
the experimental cross section 80 mb for $\pi \pi \to \rho$ \cite{FMMR}.
Assuming pointlike hadron vertices, the mesonic model in Ref. 
\cite{LK} leads to a large value of 260 mb. 

From Tables 7 and 8 we get ratios of the cross sections,
\begin{eqnarray}
& & \sigma_{D\bar{D} \to \psi (4040)}:
\sigma_{D^*\bar{D}+D\bar{D}^* \to \psi (4040)}:
\sigma_{D^*\bar{D}^* \to \psi (4040)}:
\sigma_{D_s^+D_s^- \to \psi (4040)}
\nonumber  \\
& = & 1:4.6:0.12:0.33 ,
\end{eqnarray}
\begin{eqnarray}
& & \sigma_{D\bar{D} \to \psi (4160)}:
\sigma_{D^*\bar{D}+D\bar{D}^* \to \psi (4160)}:
\sigma_{D^*\bar{D}^* \to \psi (4160)}:
\nonumber  \\
& & \sigma_{D_s^+D_s^- \to \psi (4160)}:
\sigma_{D_s^{*+}D_s^- + D_s^+D_s^{*-} \to \psi (4160)}
\nonumber  \\
& = & 1:4.2:0.17:0.15:0.14 .
\end{eqnarray}
In an analysis of the production
of charmed mesons and charmed strange mesons in the region of $\psi (4040)$ 
and $\psi (4160)$ resonances in $e^+e^-$ collisions in Ref. \cite{BIO},
the couplings of
$\psi (4040)$ and $\psi (4160)$ to two charmed mesons and charmed strange 
mesons are factorized. The couplings are assumed to be proportional to a 
coefficient (denoted as ${\rm g}_{Ri}$ in the reference), the values of which 
are listed in Table 1 of the reference.
If it is assumed that cross sections for the production of $\psi (4040)$ and 
$\psi (4160)$ from $D\bar{D}$, $D^*\bar{D}$, $D\bar{D}^*$, $D^*\bar{D}^*$, 
$D_s^+D_s^-$, $D_s^{*+}D_s^-$, $D_s^+D_s^{*-}$, and $D_s^{*+}D_s^{*-}$ 
differ from each other only by the coefficient, then ratios of the cross 
sections are,
\begin{eqnarray}
& & \sigma_{D\bar{D} \to \psi (4040)}:
\sigma_{D^*\bar{D}+D\bar{D}^* \to \psi (4040)}:
\sigma_{D^*\bar{D}^* \to \psi (4040)}:
\sigma_{D_s^+D_s^- \to \psi (4040)}
\nonumber  \\
& = & 1:4:7:1 ,
\end{eqnarray}
\begin{eqnarray}
& & \sigma_{D\bar{D} \to \psi (4160)}:
\sigma_{D^*\bar{D}+D\bar{D}^* \to \psi (4160)}:
\sigma_{D^*\bar{D}^* \to \psi (4160)}:
\nonumber  \\
& & \sigma_{D_s^+D_s^- \to \psi (4160)}:
\sigma_{D_s^{*+}D_s^- + D_s^+D_s^{*-} \to \psi (4160)}
\nonumber  \\
& = & 1:1:7.7:1:1 .
\end{eqnarray}
This assumption is correct when the $D^*$ ($\bar{D}^*$, $D_s^{*+}$, $D_s^{*-}$)
mass equals the $D$ ($\bar{D}$, $D_s^+$, $D_s^-$) mass. Regarding the ratios
given in Eq. (52) we mention a study of production of charmed mesons in 
$e^+ e^-$ collisions in Ref. \cite{RGG}. An electron and a positron annihilate
into a photon. The photon directly produces a pair of charmed quarks, each
of which becomes a charmed meson in combination with a subsequently produced
$u$ or $d$ quark. Neglecting the difference between the $D$ and $D^*$ masses
and the spin-spin interaction between the charmed quark and the light quark, 
the final states $D\bar D$, $D^*\bar{D} + D\bar{D}^*$, and $D^*\bar{D}^*$ are
populated in ratios 1:4:7, which agree with those given in Eq. (52). 
Nevertheless, the agreement does not surprise us in view of heavy quark
symmetry \cite{RGG,IW,Georgi,Neubert}. Heavy quark symmetry is that, in the
limit of very large quark mass, strong interactions of the heavy quark
become independent of its mass and spin. A consequence of heavy quark
symmetry is that $m_{D^*}=m_D$, $m_{\bar{D}^*}=m_{\bar D}$, and
$m_{D_s^{*\pm}}=m_{D_s^{\pm}}$. Therefore, heavy quark symmetry is assumed
in obtaining the ratios in Eq. (52) and in Ref. \cite{RGG}, and
the ratios 1:4:7 are general in the transition between charmed mesons and 
$c\bar c$ in heavy quark symmetry.

Cross sections for the production of $\psi(4040)$ from the fusion of two
charmed mesons are proportional to not only the couplings squared but also the
factor $\frac{1}{\sqrt{(P_A \cdot P_B)^2 -m_A^2m_B^2}}$ as seen in Eq. (29). 
When the colliding mesons go through the cases of $D\bar{D}$, $D^*\bar{D}$, and
$D^*\bar{D}^*$, the factor does not change in heavy quark symmetry, but
changes if the $D^*$ mass is not the same as the $D$ mass. The latter causes 
$\sigma_{D\bar{D} \to \psi (4040)}: \sigma_{D^*\bar{D}+D\bar{D}^* \to 
\psi (4040)}: \sigma_{D^*\bar{D}^* \to \psi (4040)}$ to deviate from the 
general ratios 1:4:7.

We have obtained cross sections for the production of $\psi (3770)$,
$\psi (4040)$, $\psi (4160)$, and $\psi (4415)$ from the fusion of two charmed
mesons and of two charmed strange mesons. As seen from the Review of Particle
Physics in Ref. \cite{PDG}, time reversal of these reactions dominate decays
of the four $c\bar c$ mesons. We study the decays in terms of the process where
a gluon is emitted by a constituent quark or antiquark in the initial mesons
and subsequently the gluon creates a quark-antiquark pair. This part is given 
in the appendix.

\vspace{0.5cm}
\leftline{\bf V. SUMMARY }
\vspace{0.5cm}

Using the process where one quark annihilates with one antiquark to create a 
gluon and subsequently the gluon is absorbed by a spectator quark or antiquark,
we have studied 2-to-1 meson-meson scattering. 
In the partial wave expansion of 
the relative-motion wave function of the two initial mesons, we have obtained
the new expressions of the transition 
amplitudes. The orbital-angular-momentum quantum number 
corresponding to the total spin of the two initial mesons is selected to
satisfy parity conservation and conservation of the total angular
momentum. The flavor and spin matrix elements have been calculated. The spin 
matrix elements corresponding to the four transition potentials are presented.
The mesonic quark-antiquark relative-motion wave functions are given by the
Schr\"odinger equation with the temperature-dependent potential. From
the transition amplitudes we have obtained the cross sections for the 
reactions: 
$K\bar{K} \to \phi$, $\pi D \to D^*$, $\pi \bar{D} \to \bar{D}^*$,
$D\bar{D} \to \psi (3770)$, $D\bar{D} \to \psi (4040)$, 
$D^*\bar{D} \to \psi (4040)$, $D\bar{D}^* \to \psi (4040)$, 
$D^*\bar{D}^* \to \psi (4040)$, $D_s^+D_s^- \to \psi (4040)$,
$D\bar{D} \to \psi (4160)$, $D^*\bar{D} \to \psi (4160)$,
$D\bar{D}^* \to \psi (4160)$, $D^*\bar{D}^* \to \psi (4160)$,
$D_s^+D_s^- \to \psi (4160)$, $D_s^{*+}D_s^- \to \psi (4160)$,
$D_s^+D_s^{*-} \to \psi (4160)$, $D\bar{D} \to \psi (4415)$,
$D^*\bar{D} \to \psi (4415)$, $D\bar{D}^* \to \psi (4415)$,
$D^*\bar{D}^* \to \psi (4415)$, $D_s^+D_s^- \to \psi (4415)$,
$D_s^{*+}D_s^- \to \psi (4415)$, $D_s^+D_s^{*-} \to \psi (4415)$, and
$D_s^{*+}D_s^{*-} \to \psi (4415)$.
The cross sections are affected by radii of initial mesons and quark masses 
that enter the transition potentials. The cross section for $K\bar{K} \to \phi$
is larger than the ones for $D\bar{D} \to \psi (3770)$, 
$D\bar{D} \to \psi (4040)$, $D\bar{D} \to \psi (4160)$, and 
$D\bar{D} \to \psi (4415)$, but smaller than the one for $\pi D \to D^*$. 
The $\psi (4040)$, $\psi (4160)$, and $\psi (4415)$ mesons resulting from 
ultrarelativistic heavy-ion collisions have been shown to be a probe
of hadronic matter that is produced in the phase transition of the quark-gluon 
plasma.

\vspace{0.5cm}
\leftline{\bf APPENDIX}
\vspace{0.5cm}

We calculate widths of the $\psi (3770)$, $\psi (4040)$, $\psi (4160)$, and 
$\psi (4415)$ decays, which produce two charmed mesons or two charmed strange 
mesons. Four Feynman diagrams at tree level are involved in $Z \to X+Y$, 
and are shown in Fig. 6. The two upper diagrams correspond to 
$Z(q_2\bar{q}_1) \to X(q_1\bar{q}_1)+Y(q_2\bar{q}_2)$, and the two lower
diagrams $Z(q_1\bar{q}_2) \to X(q_1\bar{q}_1)+Y(q_2\bar{q}_2)$. Denote by
$\psi_Z$ and $\psi_{XY}$ the mesonic quark-antiquark wave functions of meson
$Z$ and of mesons $X$ and $Y$, respectively.
Let $(E_X, \vec{p}_X)$, $(E_Y, \vec{p}_Y)$, and $(E_Z, \vec{p}_Z)$
be the four-momenta of mesons $X$, $Y$, and $Z$, respectively.
The transition amplitudes corresponding to the left upper diagram, the right
upper diagram, the left lower diagram, and the right lower diagram are given by
\begin{equation}
{\cal M}_{{\rm c}\bar{q}_1} = \sqrt {2E_Z2E_X2E_Y}
\int d\vec{r}_{q_1\bar{q}_1} d\vec{r}_{q_2\bar{q}_2}
\psi_{XY}^+ V_{{\rm c}\bar{q}_1}
\psi_Z e^{-i\vec{p}_{q_1\bar{q}_1,q_2\bar{q}_2}
\cdot\vec{r}_{q_1\bar{q}_1,q_2\bar{q}_2}},
\end{equation}
\begin{equation}
{\cal M}_{{\rm c}q_2} = \sqrt {2E_Z2E_X2E_Y}
\int d\vec{r}_{q_1\bar{q}_1} d\vec{r}_{q_2\bar{q}_2}
\psi_{XY}^+ V_{{\rm c}q_2}
\psi_Z e^{-i\vec{p}_{q_1\bar{q}_1,q_2\bar{q}_2}
\cdot\vec{r}_{q_1\bar{q}_1,q_2\bar{q}_2}},
\end{equation}
\begin{equation}
{\cal M}_{{\rm c}q_1} = \sqrt {2E_Z2E_X2E_Y}
\int d\vec{r}_{q_1\bar{q}_1} d\vec{r}_{q_2\bar{q}_2}
\psi_{XY}^+ V_{{\rm c}q_1}
\psi_Z e^{-i\vec{p}_{q_1\bar{q}_1,q_2\bar{q}_2}
\cdot\vec{r}_{q_1\bar{q}_1,q_2\bar{q}_2}},
\end{equation}
\begin{equation}
{\cal M}_{{\rm c}\bar{q}_2} = \sqrt {2E_Z2E_X2E_Y}
\int d\vec{r}_{q_1\bar{q}_1} d\vec{r}_{q_2\bar{q}_2}
\psi_{XY}^+ V_{{\rm c}\bar{q}_2}
\psi_Z e^{-i\vec{p}_{q_1\bar{q}_1,q_2\bar{q}_2}
\cdot\vec{r}_{q_1\bar{q}_1,q_2\bar{q}_2}},
\end{equation}
where $V_{{\rm c}\bar{q}_1}$, $V_{{\rm c}q_2}$, $V_{{\rm c}q_1}$, and 
$V_{{\rm c}\bar{q}_2}$ are the transition potentials for the processes where
a gluon is emitted by a constiutent quark or antiquark in the initial mesons
and subsequently the gluon creates a quark-antiquark pair, and are given in
Eqs. (51) and (52) of Ref. \cite{LXW}. The two upper diagrams give
$\bar{q}_1(p_{\bar{q}_1}^\prime) +q_2(p_{q_2}^\prime) \to
q_1(p_{q_1})+\bar{q}_1(p_{\bar{q}_1})+q_2(p_{q_2}) +\bar{q}_2(p_{\bar{q}_2})$,
and the two lower diagrams
$q_1(p_{q_1}^\prime) +\bar{q}_2(p_{\bar{q}_2}^\prime) \to 
q_1(p_{q_1})+\bar{q}_1(p_{\bar{q}_1})+q_2(p_{q_2}) +\bar{q}_2(p_{\bar{q}_2})$. 
We have
$V_{{\rm c}\bar{q}_1}=V_{{\rm r}q_1\bar{q}_2\bar{q}_1}$,
$V_{{\rm c}q_2}=V_{{\rm r}q_1\bar{q}_2q_2}$, 
$V_{{\rm c}q_1}=V_{{\rm r}q_2\bar{q}_1q_1}$, and
$V_{{\rm c}\bar{q}_2}=V_{{\rm r}q_2\bar{q}_1\bar{q}_2}$.

The transition amplitudes squared are given by
\begin{equation}
\mid {\cal M}_{{\rm c}\bar{q}_1} \mid^2 = \mid \sqrt {2E_X2E_Y2E_Z}
\int d\vec{r}_{q_1\bar{q}_1} d\vec{r}_{q_2\bar{q}_2}
\psi_Z^+ V_{{\rm c}\bar{q}_1}
\psi_{XY} e^{i\vec{p}_{q_1\bar{q}_1,q_2\bar{q}_2}
\cdot\vec{r}_{q_1\bar{q}_1,q_2\bar{q}_2}} \mid^2,
\end{equation}
\begin{equation}
\mid {\cal M}_{{\rm c}q_2} \mid^2 = \mid \sqrt {2E_X2E_Y2E_Z}
\int d\vec{r}_{q_1\bar{q}_1} d\vec{r}_{q_2\bar{q}_2}
\psi_Z^+ V_{{\rm c}q_2}
\psi_{XY} e^{i\vec{p}_{q_1\bar{q}_1,q_2\bar{q}_2}
\cdot\vec{r}_{q_1\bar{q}_1,q_2\bar{q}_2}} \mid^2,
\end{equation}
\begin{equation}
\mid {\cal M}_{{\rm c}q_1} \mid^2 = \mid \sqrt {2E_X2E_Y2E_Z}
\int d\vec{r}_{q_1\bar{q}_1} d\vec{r}_{q_2\bar{q}_2}
\psi_Z^+ V_{{\rm c}q_1}
\psi_{XY} e^{i\vec{p}_{q_1\bar{q}_1,q_2\bar{q}_2}
\cdot\vec{r}_{q_1\bar{q}_1,q_2\bar{q}_2}} \mid^2,
\end{equation}
\begin{equation}
\mid {\cal M}_{{\rm c}\bar{q}_2} \mid^2 = \mid \sqrt {2E_X2E_Y2E_Z}
\int d\vec{r}_{q_1\bar{q}_1} d\vec{r}_{q_2\bar{q}_2}
\psi_Z^+ V_{{\rm c}\bar{q}_2}
\psi_{XY} e^{i\vec{p}_{q_1\bar{q}_1,q_2\bar{q}_2}
\cdot\vec{r}_{q_1\bar{q}_1,q_2\bar{q}_2}} \mid^2.
\end{equation}
It is obvious that $\mid {\cal M}_{{\rm c}\bar{q}_1} \mid^2$, 
$\mid {\cal M}_{{\rm c}q_2} \mid^2$, $\mid {\cal M}_{{\rm c}q_1} \mid^2$, and 
$\mid {\cal M}_{{\rm c}\bar{q}_2} \mid^2$ equal 
$\mid {\cal M}_{{\rm r}q_1\bar {q}_2 \bar{q}_1} \mid^2$, 
$\mid {\cal M}_{{\rm r}q_1\bar {q}_2 q_2} \mid^2$, 
$\mid {\cal M}_{{\rm r}q_2 \bar{q}_1 q_1} \mid^2$, 
and $\mid {\cal M}_{{\rm r}q_2 \bar{q}_1 \bar{q}_2} \mid^2$, respectively.
Therefore, Eqs. (22) and (26)-(28) can be used to calculate
$\mid {\cal M}_{{\rm c}\bar{q}_1} \mid^2$, 
$\mid {\cal M}_{{\rm c}q_2} \mid^2$, $\mid {\cal M}_{{\rm c}q_1} \mid^2$, and 
$\mid {\cal M}_{{\rm c}\bar{q}_2} \mid^2$.

The transition amplitudes lead to the decay width for $Z \to X+Y$:
\begin{eqnarray}
W & = & \frac{1}{2J_Z+1}\int \frac{d^3p_X}{(2\pi)^3}\frac{d^3p_Y}{(2\pi)^3}
\frac{(2\pi)^4\delta (E_X+E_Y-E_Z)\delta^3 (\vec{p}_X+\vec{p}_Y-\vec{p}_Z)}
{2E_X2E_Y2E_Z}
\nonumber   \\
& & \sum\limits_{J_{Zz}J_{Xz}J_{Yz}}
\mid {\cal M}_{{\rm c}\bar{q}_1} +  {\cal M}_{{\rm c}q_2} 
+ {\cal M}_{{\rm c}q_1} + {\cal M}_{{\rm c}\bar{q}_2} \mid^2 .
\end{eqnarray}
where $J_i$ ($i=Z$, $X$, $Y$) is the total angular momentum of meson $i$ with
its magnetic projection quantum number $J_{iz}$.
In Table 9 decay widths are shown for the $\psi (3770)$, $\psi (4040)$, 
$\psi (4160)$, and $\psi (4415)$ decays that produce charmed mesons or
charmed strange mesons. The decay width for $\psi (3770) \to D\bar{D}$ is
18 MeV compared to the experimental value $25.3\pm 0.09$ MeV. The total widths
of the $\psi (4040)$, $\psi (4160)$, and $\psi (4415)$ decays listed in the
Review of Particle Physics \cite{PDG} are $80\pm 10$ MeV, $70\pm 10$ MeV, and
$62\pm 20$ MeV, respectively, but the partial widths for the $\psi (4040)$, 
$\psi (4160)$, and $\psi (4415)$ decays to charmed mesons or charmed strange
mesons have not been given. However, since the $\psi (4040)$, 
$\psi (4160)$, and $\psi (4415)$ decays mainly produce charmed mesons and
charmed strange mesons, the measured total widths of the $\psi (4040)$, 
$\psi (4160)$, and $\psi (4415)$ decays may be compared to 70.2 MeV as the sum
of the five $\psi (4040)$ partial widths, 63.3 MeV as the sum
of the seven $\psi (4160)$ partial widths, and 61.1 MeV as the sum
of the eight $\psi (4415)$ partial widths. We note that our partial widths
differ from those obtained in the $^3P_0$ model in Ref. \cite{BGS}.

\vspace{0.5cm}
\leftline{\bf ACKNOWLEDGEMENTS}
\vspace{0.5cm}

This work was supported by the National Natural Science Foundation of China
under Grant No. 11175111.

\newpage
\begin{figure}[htbp]
  \centering
    \includegraphics[scale=0.65]{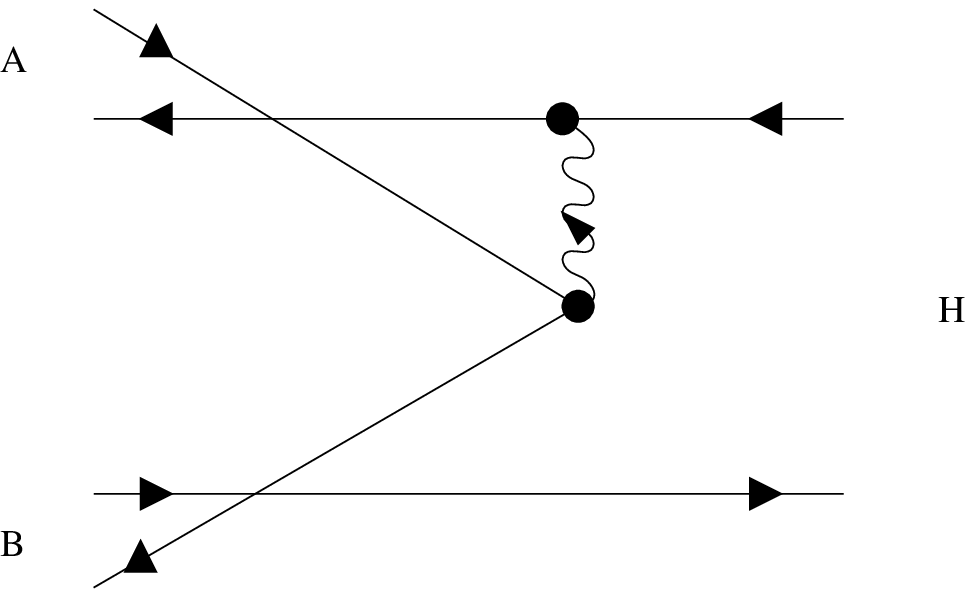}
      \hspace{2cm}
    \includegraphics[scale=0.65]{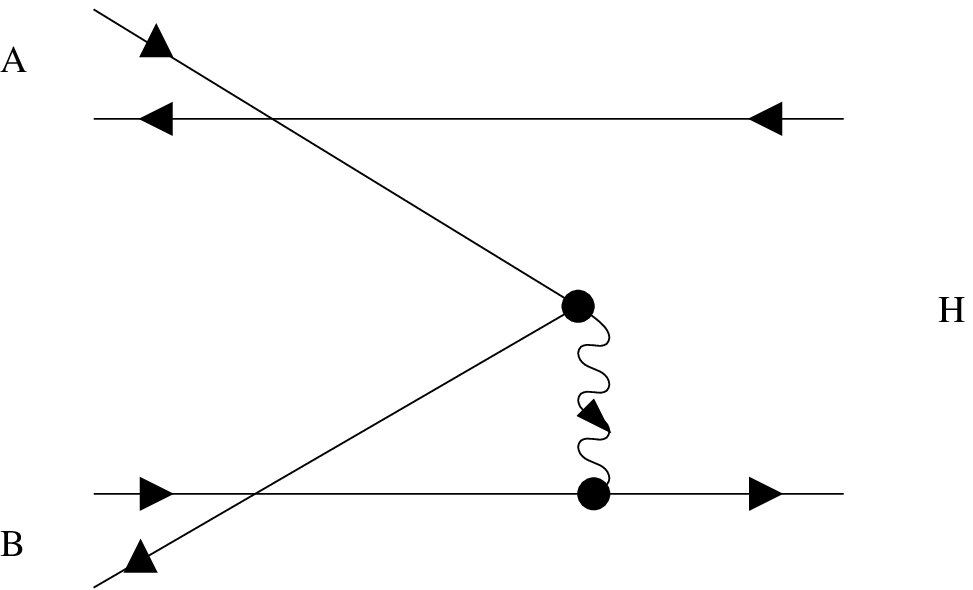}
      \vskip 72pt
    \includegraphics[scale=0.65]{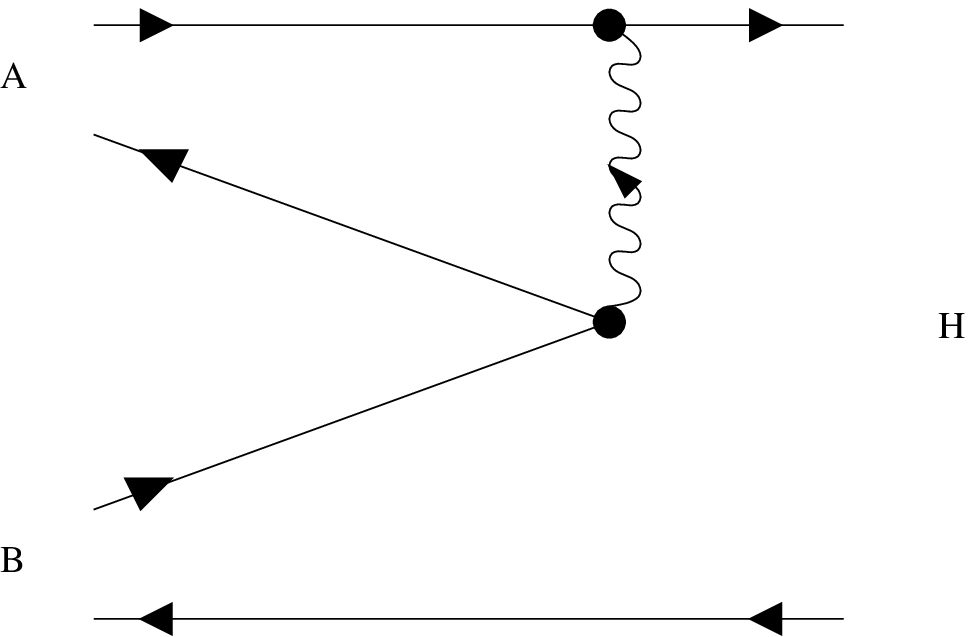}
      \hspace{2cm}
    \includegraphics[scale=0.65]{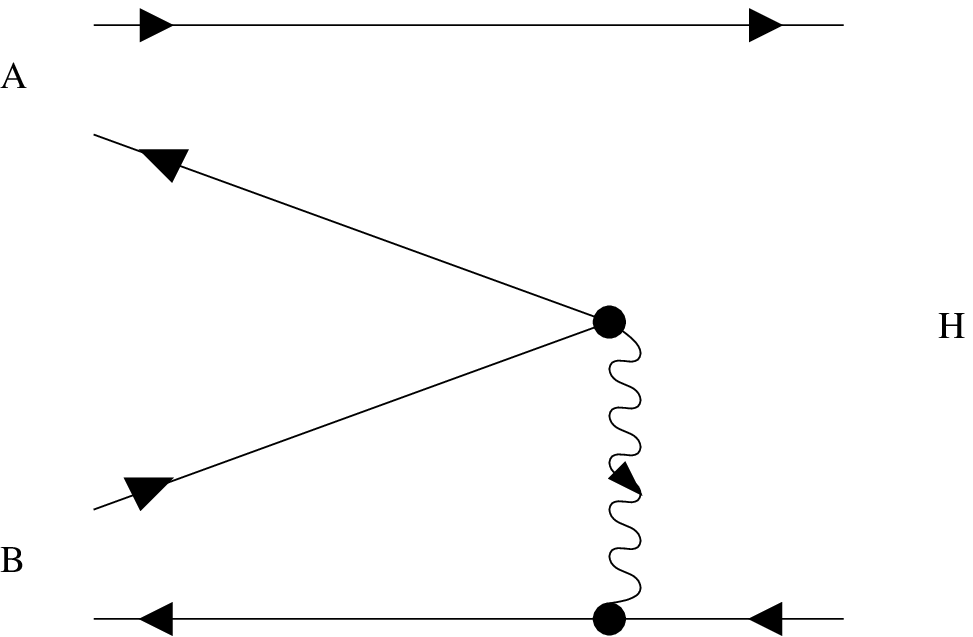}
\caption{Reaction $A+B \to H$. Solid lines with right (left) triangles stand
for quarks (antiquarks). Wavy lines stand for gluons.}
\label{fig1}
\end{figure}

\newpage
\begin{figure}[htbp]
  \centering
    \includegraphics[scale=0.65]{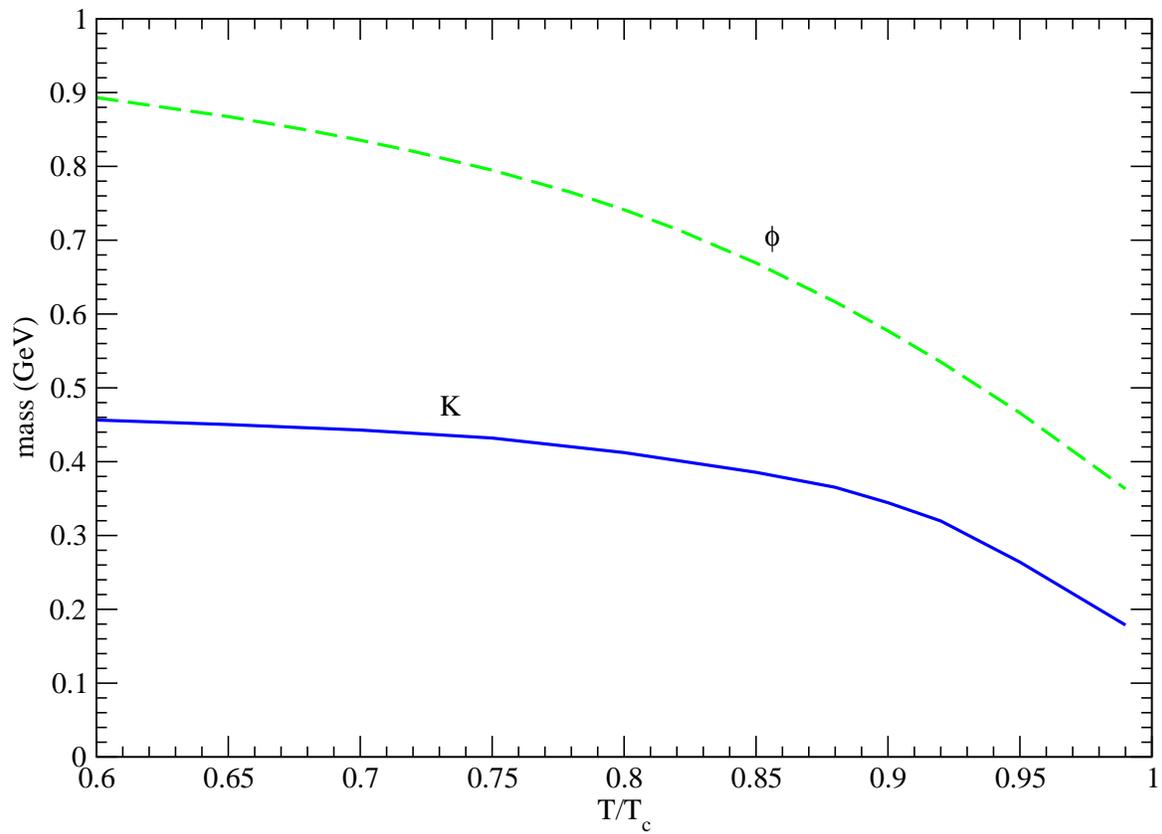}
\caption{$K$ and $\phi$ masses as functions of $T/T_{\rm c}$.}
\label{fig2}
\end{figure}

\newpage
\begin{figure}[htbp]
  \centering
    \includegraphics[scale=0.65]{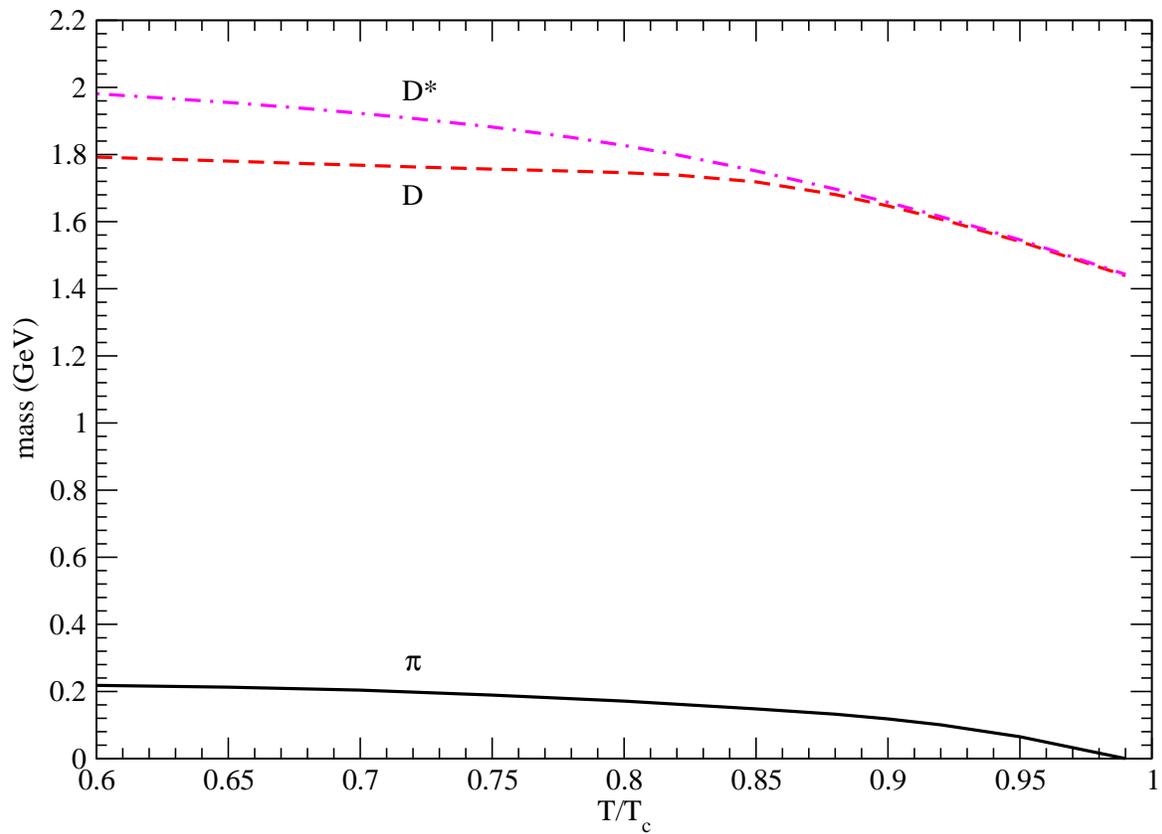}
\caption{$\pi$, $D$, and $D^\ast$ masses as functions of $T/T_{\rm c}$.}
\label{fig3}
\end{figure}

\newpage
\begin{figure}[htbp]
  \centering
    \includegraphics[scale=0.65]{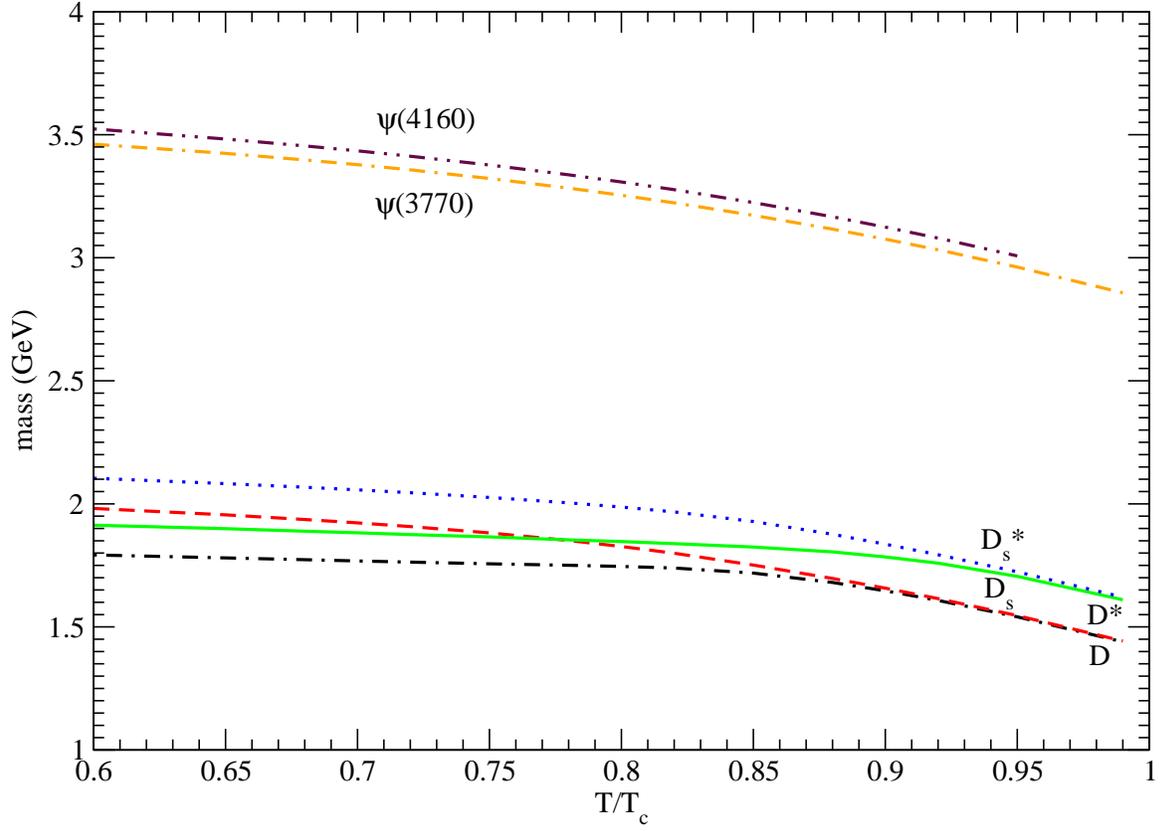}
\caption{$D$, $D^\ast$, $D_s$, $D_s^\ast$, $\psi (3770)$, and $\psi (4160)$ 
masses as functions of $T/T_{\rm c}$.}
\label{fig4}
\end{figure}

\newpage
\begin{figure}[htbp]
  \centering
    \includegraphics[scale=0.65]{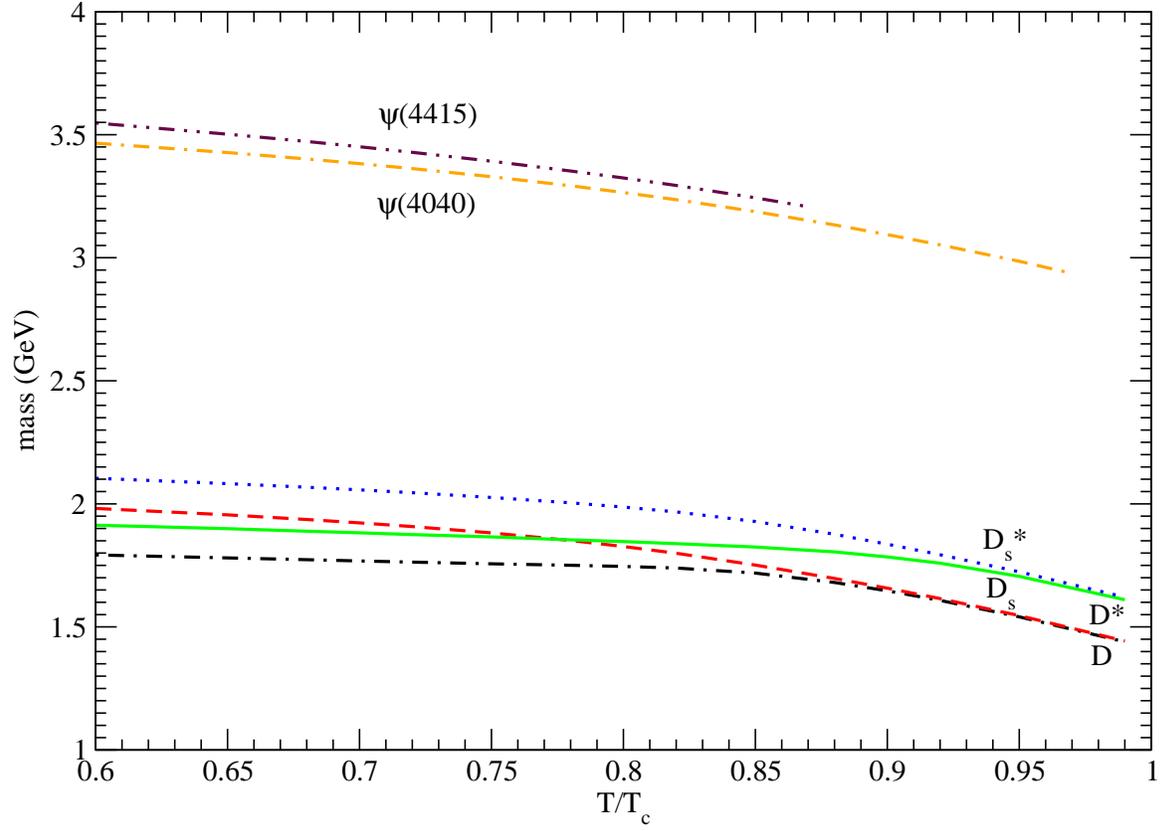}
\caption{$D$, $D^\ast$, $D_s$, $D_s^\ast$, $\psi (4040)$, and $\psi (4415)$ 
masses as functions of $T/T_{\rm c}$.}
\label{fig5}
\end{figure}

\newpage
\begin{figure}[htbp]
  \centering
    \includegraphics[scale=0.65]{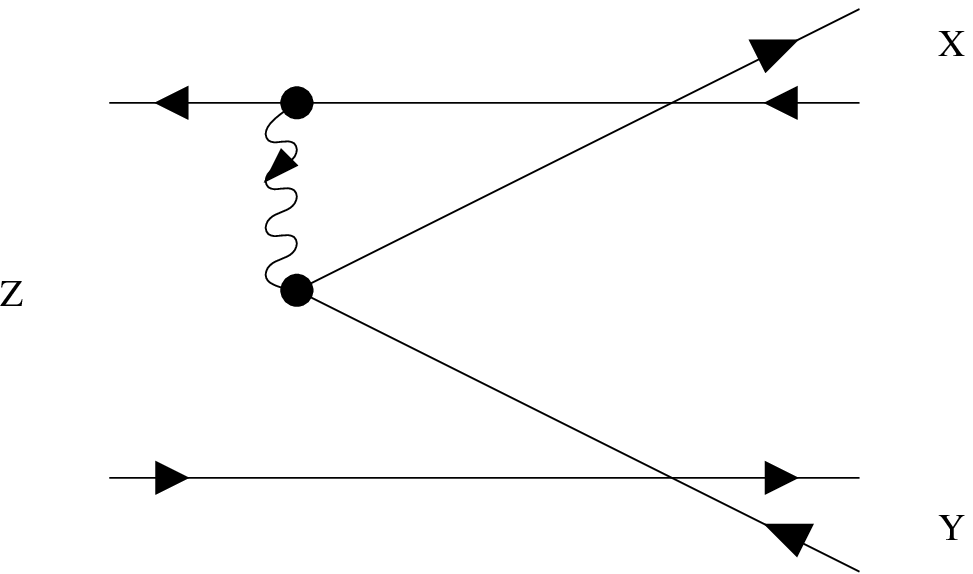}
      \hspace{2cm}
    \includegraphics[scale=0.65]{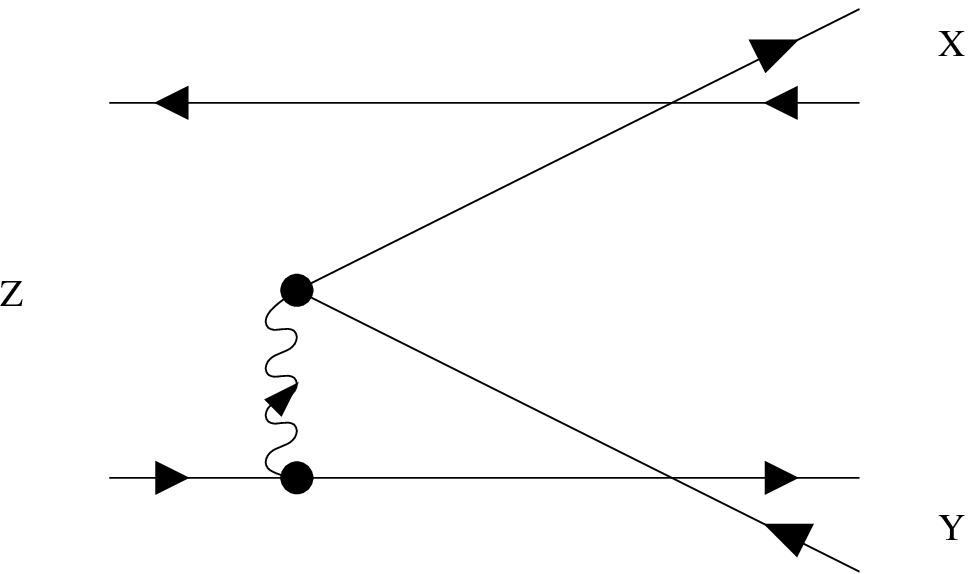}
      \vskip 72pt
    \includegraphics[scale=0.65]{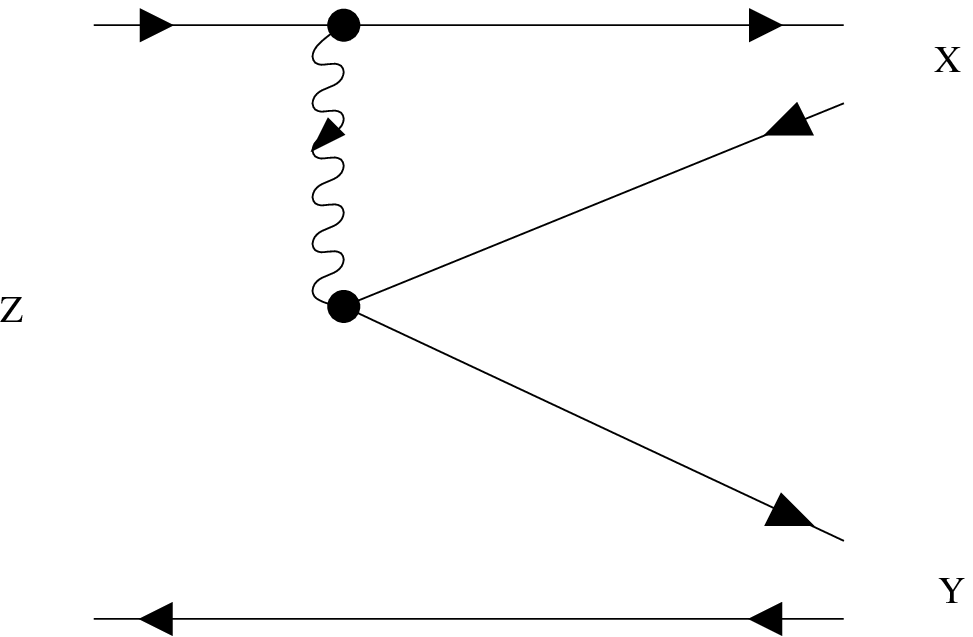}
      \hspace{2cm}
    \includegraphics[scale=0.65]{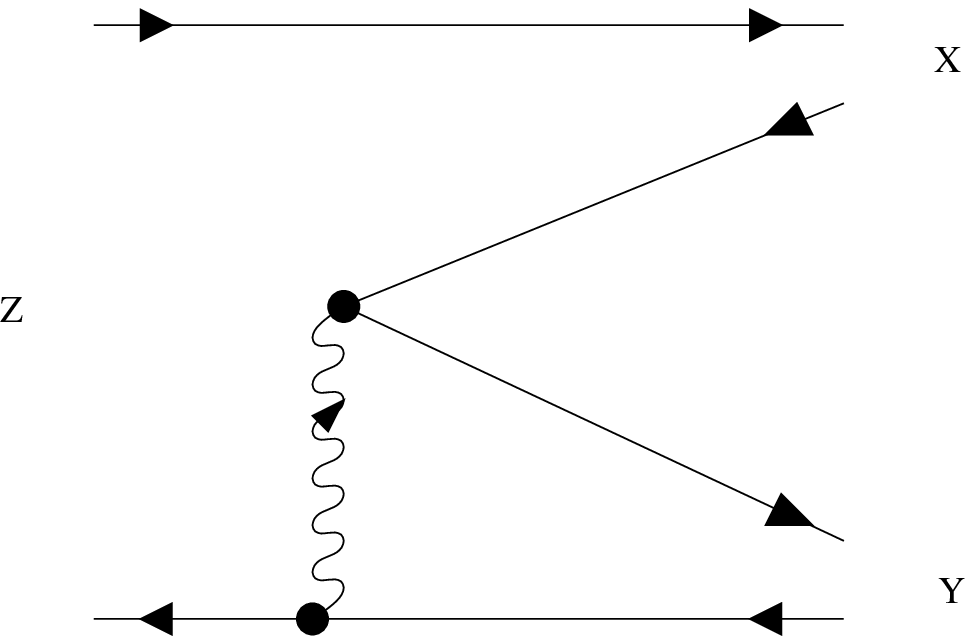}
\caption{Decay $Z \to X+Y$. Solid lines with right (left) triangles stand
for quarks (antiquarks). Wavy lines stand for gluons.}
\label{fig6}
\end{figure}

\clearpage
\begin{table*}[htbp]
\caption{\label{table1}Flavor matrix elements.}
\tabcolsep=5pt
\begin{tabular}{ccccc}
  \hline
  \hline
diagram in Fig. 1 & left upper & right upper & left lower & right lower\\
${\cal M}_{{\rm f}K\bar{K} \to \phi}$ & $\sqrt{2}$ & $\sqrt{2}$ & 0 & 0\\
${\cal M}_{{\rm f}\pi D \to D^\ast}$ & $\frac{3}{\sqrt 6}$ & 
$\frac{3}{\sqrt 6}$ & 0 & 0\\
${\cal M}_{{\rm f}D \bar{D} \to \psi (3770)}$ & 0 & 0 & -$\sqrt{2}$ & 
-$\sqrt{2}$\\
${\cal M}_{{\rm f}D_s^+ D_s^- \to \psi (4040)}$ & 0 & 0 & 1 & 1\\
  \hline
  \hline
\end{tabular}
\end{table*}

\begin{table}
\caption{\label{table2} Spin matrix elements in
${\cal M}_{{\rm r}q_1\bar{q}_2\bar{q}_1}$ for 
$A(S_A=1)+B(S_B=0) \to H(S_H=1)$.}
\begin{tabular}{cccccccccc}
\hline
$S_{Hz}$ & -1 & -1 & -1 &  0 & 0 & 0 &  1 & 1 & 1 \\
$S_{Az}$ & -1 &  0 &  1 & -1 & 0 & 1 & -1 & 0 & 1 \\
$S_{Bz}$ &  0 &  0 &  0 &  0 & 0 & 0 &  0 & 0 & 0 \\
\hline
$\phi_{\rm fss}^+ \phi_{\rm iss}$ 
& 0 & $-\frac {1}{2}$ & 0 & $\frac {1}{2}$ & 0 & $-\frac {1}{2}$ & 0 
& $\frac {1}{2}$ & 0 \\
$\phi_{\rm fss}^+ \sigma_1(21) \phi_{\rm iss}$ 
& $-\frac {1}{\sqrt 2}$ & 0 & 0 & 0 & 0 & 0 & 0 & 0 & $\frac {1}{\sqrt 2}$ \\
$\phi_{\rm fss}^+ \sigma_2(21) \phi_{\rm iss}$ 
& $\frac {i}{\sqrt 2}$ & 0 & 0 & 0 & $\frac {i}{\sqrt 2}$ & 0 & 0 & 0 
& $\frac {i}{\sqrt 2}$ \\
$\phi_{\rm fss}^+ \sigma_3(21) \phi_{\rm iss}$ 
& 0 & $-\frac {1}{2}$ & 0 & $-\frac {1}{2}$ & 0 & $-\frac {1}{2}$ & 0 
& $-\frac {1}{2}$ & 0 \\
$\phi_{\rm fss}^+ \sigma_1(1) \phi_{\rm iss}$ 
& 0 & 0 & $-\frac {1}{\sqrt 2}$ & 0 & 0 & 0 & $\frac {1}{\sqrt 2}$ & 0 & 0 \\
$\phi_{\rm fss}^+ \sigma_2(1) \phi_{\rm iss}$ 
& 0 & 0 & $\frac {i}{\sqrt 2}$ & 0 & $-\frac {i}{\sqrt 2}$ & 0 
& $\frac {i}{\sqrt 2}$ & 0 & 0 \\
$\phi_{\rm fss}^+ \sigma_3(1) \phi_{\rm iss}$ 
& 0 & $\frac {1}{2}$ & 0 & $-\frac {1}{2}$ & 0 & $-\frac {1}{2}$ & 0 
& $\frac {1}{2}$ & 0 \\
\hline
$\phi_{\rm fss}^+ \sigma_1(21) \sigma_1(1) \phi_{\rm iss}$
& 0 & $-\frac {1}{2}$ & 0 & $-\frac {1}{2}$ & 0 & $\frac {1}{2}$ & 0 
& $\frac {1}{2}$ & 0 \\
$\phi_{\rm fss}^+ \sigma_1(21) \sigma_2(1) \phi_{\rm iss}$
& 0 & $\frac {i}{2}$ & 0 & $-\frac {i}{2}$ & 0 & $-\frac {i}{2}$ & 0 
& $\frac {i}{2}$ & 0 \\
$\phi_{\rm fss}^+ \sigma_1(21) \sigma_3(1) \phi_{\rm iss}$
& $\frac {1}{\sqrt 2}$ & 0 & 0 & 0 & $-\frac {1}{\sqrt 2}$ & 0 & 0 & 0 
& $\frac {1}{\sqrt 2}$ \\
$\phi_{\rm fss}^+ \sigma_2(21) \sigma_1(1) \phi_{\rm iss}$
& 0 & $\frac {i}{2}$ & 0 & $\frac {i}{2}$ & 0 & $\frac {i}{2}$ & 0 
& $\frac {i}{2}$ & 0 \\
$\phi_{\rm fss}^+ \sigma_2(21) \sigma_2(1) \phi_{\rm iss}$
& 0 & $\frac {1}{2}$ & 0 & -$\frac {1}{2}$ & 0 & $\frac {1}{2}$ & 0 
& -$\frac {1}{2}$ & 0 \\
$\phi_{\rm fss}^+ \sigma_2(21) \sigma_3(1) \phi_{\rm iss}$
& -$\frac {i}{\sqrt 2}$ & 0 & 0 & 0 & 0 & 0 & 0 & 0 & $\frac {i}{\sqrt 2}$ \\
$\phi_{\rm fss}^+ \sigma_3(21) \sigma_1(1) \phi_{\rm iss}$
& 0 & 0 & $-\frac {1}{\sqrt 2}$ & 0 & $-\frac {1}{\sqrt 2}$ & 0 
& $-\frac {1}{\sqrt 2}$ & 0 & 0 \\
$\phi_{\rm fss}^+ \sigma_3(21) \sigma_2(1) \phi_{\rm iss}$
& 0 & 0 & $\frac {i}{\sqrt 2}$ & 0 & 0 & 0 & -$\frac {i}{\sqrt 2}$ & 0 & 0 \\
$\phi_{\rm fss}^+ \sigma_3(21) \sigma_3(1) \phi_{\rm iss}$
& 0 & $\frac {1}{2}$ & 0 & $\frac {1}{2}$ & 0 & -$\frac {1}{2}$ & 0 
& -$\frac {1}{2}$ & 0 \\
\hline
\end{tabular}
\end{table}

\begin{table}
\caption{\label{table3} Spin matrix elements in
${\cal M}_{{\rm r}q_1\bar{q}_2\bar{q}_1}$ for 
$A(S_A=0)+B(S_B=1) \to H(S_H=1)$.}
\begin{tabular}{cccccccccc}
\hline
$S_{Hz}$ & -1 & -1 & -1 &  0 & 0 & 0 &  1 & 1 & 1 \\
$S_{Az}$ &  0 &  0 &  0 &  0 & 0 & 0 &  0 & 0 & 0 \\
$S_{Bz}$ & -1 &  0 &  1 & -1 & 0 & 1 & -1 & 0 & 1 \\
\hline
$\phi_{\rm fss}^+ \phi_{\rm iss}$ 
& 0 & $\frac {1}{2}$ & 0 & -$\frac {1}{2}$ & 0 & $\frac {1}{2}$ & 0 
& -$\frac {1}{2}$ & 0 \\
$\phi_{\rm fss}^+ \sigma_1(21) \phi_{\rm iss}$ 
& $\frac {1}{\sqrt 2}$ & 0 & 0 & 0 & 0 & 0 & 0 & 0 & -$\frac {1}{\sqrt 2}$ \\
$\phi_{\rm fss}^+ \sigma_2(21) \phi_{\rm iss}$ 
& $\frac {i}{\sqrt 2}$ & 0 & 0 & 0 & $\frac {i}{\sqrt 2}$ & 0 & 0 & 0 
& $\frac {i}{\sqrt 2}$ \\
$\phi_{\rm fss}^+ \sigma_3(21) \phi_{\rm iss}$ 
& 0 & $\frac {1}{2}$ & 0 & $\frac {1}{2}$ & 0 & $\frac {1}{2}$ & 0 
& $\frac {1}{2}$ & 0 \\
$\phi_{\rm fss}^+ \sigma_1(1) \phi_{\rm iss}$ 
& $-\frac {1}{\sqrt 2}$ & 0 & 0 & 0 & 0 & 0 & 0 & 0 & $\frac {1}{\sqrt 2}$ \\
$\phi_{\rm fss}^+ \sigma_2(1) \phi_{\rm iss}$ 
& $\frac {i}{\sqrt 2}$ & 0 & 0 & 0 & $\frac {i}{\sqrt 2}$ & 0 & 0 & 0 
& $\frac {i}{\sqrt 2}$ \\
$\phi_{\rm fss}^+ \sigma_3(1) \phi_{\rm iss}$ 
& 0 & $-\frac {1}{2}$ & 0 & $-\frac {1}{2}$ & 0 & $-\frac {1}{2}$ & 0 
& $-\frac {1}{2}$ & 0 \\
\hline
$\phi_{\rm fss}^+ \sigma_1(21) \sigma_1(1) \phi_{\rm iss}$
& 0 & $-\frac {1}{2}$ & 0 & $\frac {1}{2}$ & 0 & -$\frac {1}{2}$ & 0 
& $\frac {1}{2}$ & 0 \\
$\phi_{\rm fss}^+ \sigma_1(21) \sigma_2(1) \phi_{\rm iss}$
& 0 & $\frac {i}{2}$ & 0 & $\frac {i}{2}$ & 0 & $\frac {i}{2}$ & 0 
& $\frac {i}{2}$ & 0 \\
$\phi_{\rm fss}^+ \sigma_1(21) \sigma_3(1) \phi_{\rm iss}$
& -$\frac {1}{\sqrt 2}$ & 0 & 0 & 0 & $-\frac {1}{\sqrt 2}$ & 0 & 0 & 0 
& -$\frac {1}{\sqrt 2}$ \\
$\phi_{\rm fss}^+ \sigma_2(21) \sigma_1(1) \phi_{\rm iss}$
& 0 & $\frac {i}{2}$ & 0 & $\frac {i}{2}$ & 0 & $\frac {i}{2}$ & 0 
& $\frac {i}{2}$ & 0 \\
$\phi_{\rm fss}^+ \sigma_2(21) \sigma_2(1) \phi_{\rm iss}$
& 0 & $\frac {1}{2}$ & 0 & -$\frac {1}{2}$ & 0 & $\frac {1}{2}$ & 0 
& -$\frac {1}{2}$ & 0 \\
$\phi_{\rm fss}^+ \sigma_2(21) \sigma_3(1) \phi_{\rm iss}$
& -$\frac {i}{\sqrt 2}$ & 0 & 0 & 0 & 0 & 0 & 0 & 0 & $\frac {i}{\sqrt 2}$ \\
$\phi_{\rm fss}^+ \sigma_3(21) \sigma_1(1) \phi_{\rm iss}$
& $\frac {1}{\sqrt 2}$ & 0 & 0 & 0 & $\frac {1}{\sqrt 2}$ & 0 & 0 & 0 
& $\frac {1}{\sqrt 2}$ \\
$\phi_{\rm fss}^+ \sigma_3(21) \sigma_2(1) \phi_{\rm iss}$
& -$\frac {i}{\sqrt 2}$ & 0 & 0 & 0 & 0 & 0 & 0 & 0 & $\frac {i}{\sqrt 2}$ \\
$\phi_{\rm fss}^+ \sigma_3(21) \sigma_3(1) \phi_{\rm iss}$
& 0 & -$\frac {1}{2}$ & 0 & $\frac {1}{2}$ & 0 & -$\frac {1}{2}$ & 0 
& $\frac {1}{2}$ & 0 \\
\hline
\end{tabular}
\end{table}

\begin{table}
\caption{\label{table4} Spin matrix elements in
${\cal M}_{{\rm r}q_1\bar{q}_2\bar{q}_1}$ for 
$A(S_A=1)+B(S_B=1) \to H(S_H=1)$ with $S_{Hz}=-1$.}
\begin{tabular}{cccccccccc}
\hline
$S_{Hz}$ & -1 & -1 & -1 & -1 & -1 & -1 & -1 & -1 & -1 \\
$S_{Az}$ & -1 & -1 & -1 &  0 &  0 &  0 &  1 &  1 &  1 \\
$S_{Bz}$ & -1 &  0 &  1 & -1 &  0 &  1 & -1 &  0 &  1 \\
\hline
$\phi_{\rm fss}^+ \phi_{\rm iss}$ 
& 1 & 0 & 0 & 0 & $\frac {1}{2}$ & 0 & 0 & 0 & 0 \\
$\phi_{\rm fss}^+ \sigma_1(21) \phi_{\rm iss}$ 
& 0 & $\frac {1}{\sqrt 2}$ & 0 & $\frac {1}{\sqrt 2}$ & 0 & 0 & 0 & 0 & 0 \\
$\phi_{\rm fss}^+ \sigma_2(21) \phi_{\rm iss}$ 
& 0 & -$\frac {i}{\sqrt 2}$ & 0 & $\frac {i}{\sqrt 2}$ & 0 & 0 & 0 & 0 & 0 \\
$\phi_{\rm fss}^+ \sigma_3(21) \phi_{\rm iss}$ 
& -1 & 0 & 0 & 0 & $\frac {1}{2}$ & 0 & 0 & 0 & 0 \\
$\phi_{\rm fss}^+ \sigma_1(1) \phi_{\rm iss}$ 
& 0 & 0 & 0 & $\frac {1}{\sqrt 2}$ & 0 & 0 & 0 & $\frac {1}{\sqrt 2}$ & 0 \\
$\phi_{\rm fss}^+ \sigma_2(1) \phi_{\rm iss}$ 
& 0 & 0 & 0 & -$\frac {i}{\sqrt 2}$ & 0 & 0 & 0 & $-\frac {i}{\sqrt 2}$ & 0 \\
$\phi_{\rm fss}^+ \sigma_3(1) \phi_{\rm iss}$ 
& -1 & 0 & 0 & 0 & -$\frac {1}{2}$ & 0 & 0 & 0 & 0 \\
\hline
$\phi_{\rm fss}^+ \sigma_1(21) \sigma_1(1) \phi_{\rm iss}$
& 0 & 0 & 0 & 0 & $\frac {1}{2}$ & 0 & 1 & 0 & 0 \\
$\phi_{\rm fss}^+ \sigma_1(21) \sigma_2(1) \phi_{\rm iss}$
& 0 & 0 & 0 & 0 & -$\frac {i}{2}$ & 0 & $-i$ & 0 & 0 \\
$\phi_{\rm fss}^+ \sigma_1(21) \sigma_3(1) \phi_{\rm iss}$
& 0 & -$\frac {1}{\sqrt 2}$ & 0 & $-\frac {1}{\sqrt 2}$ & 0 & 0 & 0 & 0 & 0 \\
$\phi_{\rm fss}^+ \sigma_2(21) \sigma_1(1) \phi_{\rm iss}$
& 0 & 0 & 0 & 0 & -$\frac {i}{2}$ & 0 & $i$ & 0 & 0 \\
$\phi_{\rm fss}^+ \sigma_2(21) \sigma_2(1) \phi_{\rm iss}$
& 0 & 0 & 0 & 0 & -$\frac {1}{2}$ & 0 & 1 & 0 & 0 \\
$\phi_{\rm fss}^+ \sigma_2(21) \sigma_3(1) \phi_{\rm iss}$
& 0 & $\frac {i}{\sqrt 2}$ & 0 & -$\frac {i}{\sqrt 2}$ & 0 & 0 & 0 & 0 & 0 \\
$\phi_{\rm fss}^+ \sigma_3(21) \sigma_1(1) \phi_{\rm iss}$
& 0 & 0 & 0 & $-\frac {1}{\sqrt 2}$ & 0 & 0 & 0 & $\frac {1}{\sqrt 2}$ & 0 \\
$\phi_{\rm fss}^+ \sigma_3(21) \sigma_2(1) \phi_{\rm iss}$
& 0 & 0 & 0 & $\frac {i}{\sqrt 2}$ & 0 & 0 & 0 & -$\frac {i}{\sqrt 2}$ & 0 \\
$\phi_{\rm fss}^+ \sigma_3(21) \sigma_3(1) \phi_{\rm iss}$
& 1 & 0 & 0 & 0 & -$\frac {1}{2}$ & 0 & 0 & 0 & 0 \\
\hline
\end{tabular}
\end{table}

\begin{table}
\caption{\label{table5} Spin matrix elements in
${\cal M}_{{\rm r}q_1\bar{q}_2\bar{q}_1}$ for 
$A(S_A=1)+B(S_B=1) \to H(S_H=1)$ with $S_{Hz}=0$.}
\begin{tabular}{cccccccccc}
\hline
$S_{Hz}$ &  0 &  0 &  0 &  0 & 0 & 0 &  0 & 0 & 0 \\
$S_{Az}$ & -1 & -1 & -1 &  0 & 0 & 0 &  1 & 1 & 1 \\
$S_{Bz}$ & -1 &  0 &  1 & -1 & 0 & 1 & -1 & 0 & 1 \\
\hline
$\phi_{\rm fss}^+ \phi_{\rm iss}$ 
& 0 & $\frac {1}{2}$ & 0 & $\frac {1}{2}$ & 0 & $\frac {1}{2}$ & 0 
& $\frac {1}{2}$ & 0 \\
$\phi_{\rm fss}^+ \sigma_1(21) \phi_{\rm iss}$ 
& 0 & 0 & $\frac {1}{\sqrt 2}$ & 0 & $\frac {1}{\sqrt 2}$ & 0 
& $\frac {1}{\sqrt 2}$ & 0 & 0 \\
$\phi_{\rm fss}^+ \sigma_2(21) \phi_{\rm iss}$ 
& 0 & 0 & -$\frac {i}{\sqrt 2}$ & 0 & 0 & 0 & $\frac {i}{\sqrt 2}$ & 0 & 0 \\
$\phi_{\rm fss}^+ \sigma_3(21) \phi_{\rm iss}$ 
& 0 & $-\frac {1}{2}$ & 0 & $-\frac {1}{2}$ & 0 & $\frac {1}{2}$ & 0 
& $\frac {1}{2}$ & 0 \\
$\phi_{\rm fss}^+ \sigma_1(1) \phi_{\rm iss}$ 
& $\frac {1}{\sqrt 2}$ & 0 & 0 & 0 & $\frac {1}{\sqrt 2}$ & 0 & 0 & 0 
& $\frac {1}{\sqrt 2}$\\
$\phi_{\rm fss}^+ \sigma_2(1) \phi_{\rm iss}$ 
& $\frac {i}{\sqrt 2}$ & 0 & 0 & 0 & 0 & 0 & 0 & 0 & -$\frac {i}{\sqrt 2}$ \\
$\phi_{\rm fss}^+ \sigma_3(1) \phi_{\rm iss}$ 
& 0 & -$\frac {1}{2}$ & 0 & $\frac {1}{2}$ & 0 & $-\frac {1}{2}$ & 0 
& $\frac {1}{2}$ & 0 \\
\hline
$\phi_{\rm fss}^+ \sigma_1(21) \sigma_1(1) \phi_{\rm iss}$
& 0 & $\frac {1}{2}$ & 0 & $\frac {1}{2}$ & 0 & $\frac {1}{2}$ & 0 
& $\frac {1}{2}$ & 0 \\
$\phi_{\rm fss}^+ \sigma_1(21) \sigma_2(1) \phi_{\rm iss}$
& 0 & $\frac {i}{2}$ & 0 & $\frac {i}{2}$ & 0 & $-\frac {i}{2}$ & 0 
& -$\frac {i}{2}$ & 0 \\
$\phi_{\rm fss}^+ \sigma_1(21) \sigma_3(1) \phi_{\rm iss}$
& 0 & 0 & -$\frac {1}{\sqrt 2}$ & 0 & 0 & 0 & $\frac {1}{\sqrt 2}$ & 0 & 0 \\
$\phi_{\rm fss}^+ \sigma_2(21) \sigma_1(1) \phi_{\rm iss}$
& 0 & -$\frac {i}{2}$ & 0 & $\frac {i}{2}$ & 0 & -$\frac {i}{2}$ & 0 
& $\frac {i}{2}$ & 0 \\
$\phi_{\rm fss}^+ \sigma_2(21) \sigma_2(1) \phi_{\rm iss}$
& 0 & $\frac {1}{2}$ & 0 & -$\frac {1}{2}$ & 0 & -$\frac {1}{2}$ & 0 
& $\frac {1}{2}$ & 0 \\
$\phi_{\rm fss}^+ \sigma_2(21) \sigma_3(1) \phi_{\rm iss}$
& 0 & 0 & $\frac {i}{\sqrt 2}$ & 0 & -$\frac {i}{\sqrt 2}$ & 0 
& $\frac {i}{\sqrt 2}$ & 0 & 0 \\
$\phi_{\rm fss}^+ \sigma_3(21) \sigma_1(1) \phi_{\rm iss}$
& $-\frac {1}{\sqrt 2}$ & 0 & 0 & 0 & 0 & 0 & 0 & 0 & $\frac {1}{\sqrt 2}$ \\
$\phi_{\rm fss}^+ \sigma_3(21) \sigma_2(1) \phi_{\rm iss}$
& -$\frac {i}{\sqrt 2}$ & 0 & 0 & 0 & $\frac {i}{\sqrt 2}$ & 0 & 0 & 0 
& -$\frac {i}{\sqrt 2}$\\
$\phi_{\rm fss}^+ \sigma_3(21) \sigma_3(1) \phi_{\rm iss}$
& 0 & $\frac {1}{2}$ & 0 & -$\frac {1}{2}$ & 0 & -$\frac {1}{2}$ & 0 
& $\frac {1}{2}$ & 0 \\
\hline
\end{tabular}
\end{table}

\begin{table}
\caption{\label{table6} Spin matrix elements in
${\cal M}_{{\rm r}q_1\bar{q}_2\bar{q}_1}$ for 
$A(S_A=1)+B(S_B=1) \to H(S_H=1)$ with $S_{Hz}=1$.}
\begin{tabular}{cccccccccc}
\hline
$S_{Hz}$ &  1 &  1 &  1 &  1 & 1 & 1 &  1 & 1 & 1 \\
$S_{Az}$ & -1 & -1 & -1 &  0 & 0 & 0 &  1 & 1 & 1 \\
$S_{Bz}$ & -1 &  0 &  1 & -1 & 0 & 1 & -1 & 0 & 1 \\
\hline
$\phi_{\rm fss}^+ \phi_{\rm iss}$ 
& 0 & 0 & 0 & 0 & $\frac {1}{2}$ & 0 & 0 & 0 & 1 \\
$\phi_{\rm fss}^+ \sigma_1(21) \phi_{\rm iss}$ 
& 0 & 0 & 0 & 0 & 0 & $\frac {1}{\sqrt 2}$ & 0 & $\frac {1}{\sqrt 2}$ & 0 \\
$\phi_{\rm fss}^+ \sigma_2(21) \phi_{\rm iss}$ 
& 0 & 0 & 0 & 0 & 0 & -$\frac {i}{\sqrt 2}$ & 0 & $\frac {i}{\sqrt 2}$ & 0 \\
$\phi_{\rm fss}^+ \sigma_3(21) \phi_{\rm iss}$ 
& 0 & 0 & 0 & 0 & $-\frac {1}{2}$ & 0 & 0 & 0 & 1 \\
$\phi_{\rm fss}^+ \sigma_1(1) \phi_{\rm iss}$ 
& 0 & $\frac {1}{\sqrt 2}$ & 0 & 0 & 0 & $\frac {1}{\sqrt 2}$ & 0 & 0 & 0 \\
$\phi_{\rm fss}^+ \sigma_2(1) \phi_{\rm iss}$ 
& 0 & $\frac {i}{\sqrt 2}$ & 0 & 0 & 0 & $\frac {i}{\sqrt 2}$ & 0 & 0 & 0 \\
$\phi_{\rm fss}^+ \sigma_3(1) \phi_{\rm iss}$ 
& 0 & 0 & 0 & 0 & $\frac {1}{2}$ & 0 & 0 & 0 & 1 \\
\hline
$\phi_{\rm fss}^+ \sigma_1(21) \sigma_1(1) \phi_{\rm iss}$
& 0 & 0 & 1 & 0 & $\frac {1}{2}$ & 0 & 0 & 0 & 0 \\
$\phi_{\rm fss}^+ \sigma_1(21) \sigma_2(1) \phi_{\rm iss}$
& 0 & 0 & $i$ & 0 & $\frac {i}{2}$ & 0 & 0 & 0 & 0 \\
$\phi_{\rm fss}^+ \sigma_1(21) \sigma_3(1) \phi_{\rm iss}$
& 0 & 0 & 0 & 0 & 0 & $\frac {1}{\sqrt 2}$ & 0 & $\frac {1}{\sqrt 2}$ & 0 \\
$\phi_{\rm fss}^+ \sigma_2(21) \sigma_1(1) \phi_{\rm iss}$
& 0 & 0 & -$i$ & 0 & $\frac {i}{2}$ & 0 & 0 & 0 & 0 \\
$\phi_{\rm fss}^+ \sigma_2(21) \sigma_2(1) \phi_{\rm iss}$
& 0 & 0 & 1 & 0 & -$\frac {1}{2}$ & 0 & 0 & 0 & 0 \\
$\phi_{\rm fss}^+ \sigma_2(21) \sigma_3(1) \phi_{\rm iss}$
& 0 & 0 & 0 & 0 & 0 & -$\frac {i}{\sqrt 2}$ & 0 & $\frac {i}{\sqrt 2}$ & 0 \\
$\phi_{\rm fss}^+ \sigma_3(21) \sigma_1(1) \phi_{\rm iss}$
& 0 & $-\frac {1}{\sqrt 2}$ & 0 & 0 & 0 & $\frac {1}{\sqrt 2}$ & 0 & 0 & 0 \\
$\phi_{\rm fss}^+ \sigma_3(21) \sigma_2(1) \phi_{\rm iss}$
& 0 & -$\frac {i}{\sqrt 2}$ & 0 & 0 & 0 & $\frac {i}{\sqrt 2}$ & 0 & 0 & 0 \\
$\phi_{\rm fss}^+ \sigma_3(21) \sigma_3(1) \phi_{\rm iss}$
& 0 & 0 & 0 & 0 & -$\frac {1}{2}$ & 0 & 0 & 0 & 1 \\
\hline
\end{tabular}
\end{table}

\begin{table*}[htbp]
\caption{\label{table7} Total spin,
orbital-angular-momentum quantum number, and cross section.}
\tabcolsep=5pt
\begin{tabular}{cccc}
  \hline
  \hline
reaction & $S$ & $L_{\rm i}$ & $\sigma^{\rm unpol}$ (mb)\\
  \hline
$K\bar{K} \to \phi$ & 0 & 1 & 8.05\\
$\pi D \to D^\ast$ & 0 & 1 & 40.21\\
$\pi \bar{D} \to \bar{D}^\ast$ & 0 & 1 & 40.21\\
  \hline
$D \bar{D} \to \psi (3770)$ & 0 & 1 & 4.28\\
  \hline
$D \bar{D} \to \psi (4040)$ & 0 & 1 & 3.45\\
$D^* \bar{D} \to \psi (4040)$ & 1 & 1 & 7.9\\
$D \bar{D}^* \to \psi (4040)$ & 1 & 1 & 7.9\\
  \hline
$D^* \bar{D}^* \to \psi (4040)$ & 0 & 1 & 0.42\\
        & 1 & 1   & \\
        & 2 & 1,3 & \\
  \hline
$D_s^+ D_s^- \to \psi (4040)$ & 0 & 1 & 1.13\\
  \hline
  \hline
\end{tabular}
\end{table*}

\begin{table*}[htbp]
\caption{\label{table8} The same as Table 7.}
\tabcolsep=5pt
\begin{tabular}{cccc}
  \hline
  \hline
reaction & $S$ & $L_{\rm i}$ & $\sigma^{\rm unpol}$ (mb)\\
  \hline
$D \bar{D} \to \psi (4160)$ & 0 & 1 & 3.35\\
$D^* \bar{D} \to \psi (4160)$ & 1 & 1 & 7.06\\
$D \bar{D}^* \to \psi (4160)$ & 1 & 1 & 7.06\\
  \hline
$D^* \bar{D}^* \to \psi (4160)$ & 0 & 1 & 0.57\\
        & 1 & 1   & \\
        & 2 & 1,3 & \\
  \hline
$D_s^+ D_s^- \to \psi (4160)$ & 0 & 1 & 0.5\\
$D_s^{*+} D_s^- \to \psi (4160)$ & 1 & 1 & 0.23\\
$D_s^+ D_s^{*-} \to \psi (4160)$ & 1 & 1 & 0.23\\
  \hline
$D \bar{D} \to \psi (4415)$ & 0 & 1 & 0.35\\
$D^* \bar{D} \to \psi (4415)$ & 1 & 1 & 5.46\\
$D \bar{D}^* \to \psi (4415)$ & 1 & 1 & 5.46\\
  \hline
$D^* \bar{D}^* \to \psi (4415)$ & 0 & 1 & 1.39\\
        & 1 & 1   & \\
        & 2 & 1,3 & \\
  \hline
$D_s^+ D_s^- \to \psi (4415)$ & 0 & 1 & 0.13\\
$D_s^{*+} D_s^- \to \psi (4415)$ & 1 & 1 & 0.75\\
$D_s^+ D_s^{*-} \to \psi (4415)$ & 1 & 1 & 0.75\\
  \hline
$D_s^{*+} D_s^{*-} \to \psi (4415)$ & 0 & 1 & 0.11\\
        & 1 & 1   & \\
        & 2 & 1,3 & \\
  \hline
  \hline
\end{tabular}
\end{table*}

\begin{table}
\centering \caption{\label{table9} Decay modes and decay widths of
$\psi (3770)$, $\psi (4040)$, and $\psi (4160)$.}
\begin{tabular}{|l|l|l|l|}
  \hline
  decay mode & decay width (MeV) & decay mode & decay width (MeV) \\
  \hline
  $\psi (3770) \to D\bar{D}$  & 18  
& $\psi (4160) \to D\bar{D}$  & 22.5 \\
  \hline
  $\psi (4040) \to D\bar{D}$    & 21.7  
& $\psi (4160) \to D^*\bar{D}$  & 4.4 \\
  \hline
  $\psi (4040) \to D^*\bar{D}$    & 22.8  
& $\psi (4160) \to D\bar{D}^*$    & 4.4 \\
  \hline
  $\psi (4040) \to D\bar{D}^*$    & 22.8  
& $\psi (4160) \to D^*\bar{D}^*$  & 29.4 \\
  \hline
  $\psi (4040) \to D^*\bar{D}^*$  & 0.52   
& $\psi (4160) \to D_s^+D_s^-$    & 2.2 \\
  \hline
  $\psi (4040) \to D_s^+D_s^-$    & 2.4    
& $\psi (4160) \to D_s^{*+}D_s^-$ & 0.22 \\
  \hline
  five $\psi (4040)$ modes        & 70.2  
& $\psi (4160) \to D_s^+D_s^{*-}$ & 0.22 \\
  \hline
                             & 
& seven $\psi (4160)$ modes  & 63.3 \\
  \hline
  \end{tabular}
\end{table}

\begin{table}
\centering \caption{\label{table10} $\psi (4415)$ decay modes and decay 
widths.}
\begin{tabular}{|l|l|}
  \hline
  decay mode & decay width (MeV) \\
  \hline
  $\psi (4415) \to D\bar{D}$  & 20.4 \\
  \hline
  $\psi (4415) \to D^*\bar{D}$  & 10.9 \\
  \hline
  $\psi (4415) \to D\bar{D}^*$  & 10.9 \\
  \hline
  $\psi (4415) \to D^*\bar{D}^*$  & 2.4 \\
  \hline
  $\psi (4415) \to D_s^+D_s^-$  & 2.4 \\
  \hline
  $\psi (4415) \to D_s^{*+}D_s^-$  & 3.5 \\
  \hline
  $\psi (4415) \to D_s^+D_s^{*-}$  & 3.5 \\
  \hline
  $\psi (4415) \to D_s^{*+}D_s^{*-}$  & 7 \\
  \hline
  eight $\psi (4415)$ modes  & 61 \\
  \hline
  \end{tabular}
\end{table}

\end{document}